# SOLPS-ITER Numerical Simulations of ITER-scale Snowflake Divertors: Low-Field-Side SF-/SF+ and High-Field-Side SF-/SF+ Configurations

H.S. Wu[1], F. Subba[1], M.R.K. Wigram[2], O. Pan[3], R. Lo Frano[4], A. Pucciarelli[4] and R. Zanino[1]

1. NEMO Group, Dipartimento Energia, Politecnico di Torino, Corso Duca degli Abruzzi 24, 10129 Torino, Italy.
2. MIT Plasma Science and Fusion Center, Cambridge, MA 02139, USA
3. Max-Planck-Institut für Plasmaphysik, 85748 Garching, Germany
4. DICI, University of Pisa, lg. L. Lazzarino n.1 Pisa, Italy

E-mail: haosheng.wu@polito.it

## Abstract

With edge plasma code SOLS-ITER, we study four Snowflake (SF) configurations for an ITER-size tokamak, with toroidal magnetic field $B_T$~5T, major radius $R$~5m and plasma current $I_p$~10MA. Our aim is to provide insights on SF divertor design for future devices. In this work, the impacts of magnetic geometry and divertor target geometry in the four types of SF configurations on plasma behavior and power exhaust performance are investigated in detail. Low-recycling regime, high-recycling and detachment in the four types of SF divertors are obtained through an upstream density scan. The secondary X-point positions of SF divertors are systematically varied to examine their impact. For Low-Field-Side (LFS) SF- and High-Field-Side (HFS) SF- divertors the observed power splitting, induced by the secondary X-point, is consistent with experimental observations. The effect of target geometry is studied by comparing the flat target plates with the ITER-like divertor shape. The overall simulation results reveal a notable consequence of the LFS SF- divertor: closed structure of the inner target with high inclined plate can compress recycling neutrals originating from the HFS divertor region towards into the LFS SOL and PFR regions. This results in considerable volumetric dissipation through strong ionization and recombination, causing the connected outer target region to detach. This feature can be considered in the design of the LFS SF− divertor for future devices. For the LFS and HFS SF+ divertors, the region between the two X-points exhibits strong ionization that recombination sources and are close to the primary X-point. This feature might be beneficial for the formation of X-point radiator, but would require further impurity seeding simulation study.

## 1. Introduction

For nuclear fusion reactors, e.g. the EU-DEMO [1], J-DEMO [2] and ARC [3], a significant amount of power will be exhausted into the Scape-Off Layer (SOL) region. In steady state, the parallel heat flux at the divertor targets, without any mitigation, can be several hundreds of $MWm^{-2}$ or even serval $GWm^{-2}$ [4]. This value far exceeds the current engineering limitation ~10-20 $MWm^{-2}$ leading to what is known as the power exhaust (PEX) problem.

We briefly describe a model that illustrates the PEX problem physics. A magnetic flux tube is considered, from upstream to the target. The perpendicular heat flux at target $q_{\perp,Tgt}$ is determined by parallel heat flux at target $q_{\parallel,Tgt}$, pitch angle $(\frac{B_p}{B})_{Tgt}$ and poloidal tilting angle $\alpha$ as in (1). $q_{\parallel,Tgt}$ depends on the upstream parallel heat flux $q_{\parallel,Up}$, the ratio of the major radius at the upstream to that at the target $\frac{R_{Up}}{R_{Tgt}}$, and energy dissipation due to volumetric process which can be represented as $f_{diss}$ in (2). The $f_{diss}$ involve radiation, ionization, dissociation of recycling atoms and molecules, recombination, and perpendicular transport. $q_{\parallel,Up}$ can be estimated by the power entering SOL region $P_{SOL}$, and the effective area $A_{\parallel,Up}$ which can be expressed using the pitch angle $(\frac{B_p}{B})_{Up}$, the upstream major radius $R_{up}$, and the power width $\lambda_q$ as detailed in equation (3) below.

$$q_{\perp,Tgt} = q_{\parallel,Tgt}(\frac{B_p}{B})_{Tgt}\sin\alpha \qquad (1)$$

$$q_{\parallel,Tgt} = q_{\parallel,Up}\frac{R_{Up}}{R_{Tgt}}(1-f_{diss}) \qquad (2)$$

$$q_{\parallel,Up} = \frac{P_{SOL}}{A_{\parallel,Up}} = \frac{P_{SOL}}{2\pi R_{Up} R(\frac{B_p}{B})_{Up}\lambda_q} \qquad (3)$$

By combining (1)-(3), it is easy to see that the $q_{\perp,Tgt}$ is affected by major radius $R_{tgt}$, flux expansion $f_x$, poloidal tilting angle $\alpha$ and volumetric dissipation $f_{diss}$ as shown in (4). The target surface heat load $q_{surf}$ includes plasma contribution $q_{\perp,Tgt}$ and the recombination which takes place in the surface with radiation effects neglected. The recombination contribution can be estimated by the ion flux $\Gamma$ and potential energy $E_{pot}$.

$$q_{\perp,Tgt} = \frac{P_{SOL}}{2\pi R_{tgt}\lambda_q}\frac{(\frac{B_p}{B})_{Tgt}}{(\frac{B_p}{B})_{Up}}\sin\alpha(1-f_{diss}) = \frac{P_{SOL}}{2\pi R_{tgt}\lambda_q}\frac{1}{f_x}\sin\alpha(1-f_{diss}) \qquad (4)$$

$$q_{surf} = q_{\perp,Tgt} + \Gamma E_{pot} \qquad (5)$$

Alternative (Advanced) divertor configurations (ADCs) [5] are investigated as the potential solutions for the PEX problem. These concepts rely on varying parameters $R_{tgt}$, $f_x$, $\sin\alpha$ and $f_{diss}$ to reduce $q_{surf}$ as much as possible. One of ADCs is the Snowflake (SF) divertor [6]. In its idealized form, this features a second-order null point. In real experimental discharges, this is realized by creating two separate X-points, close to each other. Compared to the conventional Single Null (SN) divertor, the two X-points in the snowflake (SF) configuration create a region with a weak poloidal magnetic field, resulting in an increased connection length and an expanded SOL volume [5]. These features can facilitate volumetric dissipation and benefit detachment.

From a topology perspective, according to the position of the secondary X-point relative to the primary one, the snowflake configuration is classified into four types [7]: Low-Field-Side Snowflake minus (LFS SF-), Low-Field-Side Snowflake plus (LFS SF+), High-Field-Side Snowflake minus (HFS SF-) and High-Field-Side Snowflake plus (HFS SF+), as shown in Figure *1*. For the LFS SF– topology, when the distance between the primary and secondary X-points is significantly extended, the configuration is referred to as an X-point Target (XPT) divertor [8][9]. Its relationship with the standard LFS SF- is analogous to the relationship between the Super-X divertor and the conventional SN divertor.

In recent years, the SF divertor has been investigated through both experiments and simulations. Experimentally, SF configurations have been implemented in devices such as TCV [10][11] [13], MAST-U [14] and DIII-D [15] confirming that the heat load on the divertor targets are reduced. Also, various edge plasma codes have been employed to study SOL plasma transport in LFS SF divertors. The state-of-the-art SOLPS-ITER code package [16], particularly the B2.5-EIRENE solver [17] combined with external mesh generators, has been applied to the upper LFS SF- divertor of ASDEX Upgrade (AUG) [18] and the LFS SF divertors in MAST-U [19]. In AUG simulations [18], the $SF^-$ divertor is found to enhance radiation compared with the standard SN case under similar conditions. Other edge plasma codes have also been utilized, including UEDGE for MAST-U [20], SOLEDGE for HL-2M [21], and EMC3-EIRENE for TCV [22] and AUG [23].

The effects of magnetic and divertor geometry in the SN divertor have been well studied [24], guiding the design of the ITER SN divertor [25]. In contrast, the effects in SF divertors remain less explored. This represents a significant research gap, especially given the growing interest in employing this configuration in future devices such as DTT [26], EU-DEMO [5], and J-DEMO [2]. To bridge this gap, we systematically investigate the impact of magnetic geometry and divertor geometry across the four types of SF configurations on plasma behavior and power exhaust performance, aiming to support the design of future devices. To this end, we simulate the four types of SF divertors at the ITER scale. This also includes, for the first time, simulations of HFS SF-/SF+ divertors in SOLPS-ITER.

This paper is structured as follows: Section 2 introduces the snowflake equilibria used in this work. Section 3 outlines the SOLPS-ITER modeling setup. Section 4 presents a comparative analysis between the four configurations. Section 5 discusses the magnetic geometry effect through the scan of the secondary X-point position in each SF configurations. Section 6 examines the impact of target geometry by comparing a flat divertor shape with an ITER-like divertor shape. The summary and outlook are provided in Section 7.

## 2. Equilibria

### 2.1. Reference equilibria

The equilibria used in this work are generated using FreeGS [27], which is a free-boundary tokamak equilibrium solver. The conventional SN equilibrium and the four types of snowflake equilibria are shown in Figure 1, named SN Reference, LFS SF− Reference, LFS SF+ Reference, HFS SF+ Reference and HFS SF− Reference. They will be used for SOLPS-ITER simulations in Section 4 and Section 6. In this study, due to the large number of plots across different configurations, a consistent color scheme is adopted in the remaining: blue for the LFS SF– configuration, orange for LFS SF+, green for HFS SF+, and purple for HFS SF–. With the same divertor geometry, the SN reference is used as a baseline for evaluating the performance of the snowflake divertors. The main parameters are comparable to ITER [28] [29]: major radius $R$~5m, minor radius $a$~1.7m, toroidal magnetic field $B_T$~5T and plasma current $I_p$~10MA, as summarized in Table 1. The magnetic geometry of LFS and HFS SF- equilibria can be characterized [11][30] by: *dxx* which is the distance between the two X-points measured at the Outer Midplane (OMP); normalized distance *σ* which is the spatial distance between the Two X-points in the poloidal plane divided by the plasma minor radius; and *θ* which is the angle between a line connecting the two X-points and a line perpendicular to the segment connecting the magnetic axis and the primary X-point. For the SF+ equilibria, there are only *σ* and *θ* because the secondary X-point is in the Private Flux Region (PFR) and cannot be directly mapped to the OMP along a magnetic field line. These magnetic geometry values for the reference equilibria are also summarized in Table 1.

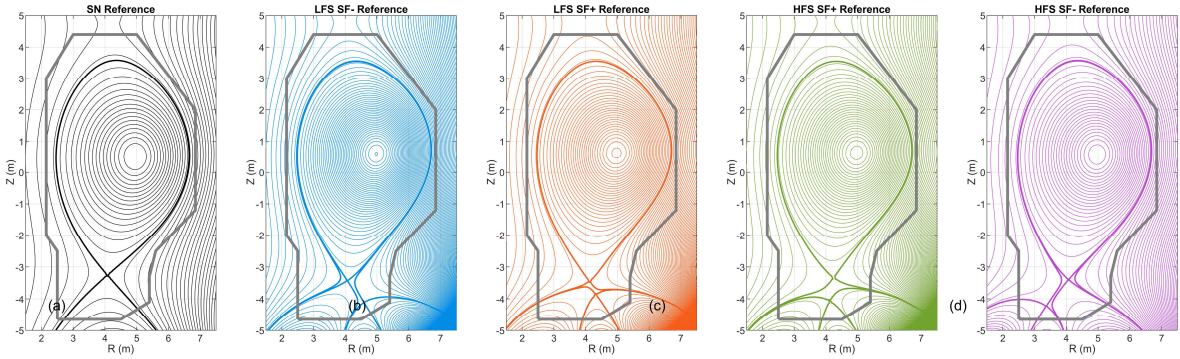

*Figure 1 The reference equilibria: (a) Conventional Single Null (SN) configuration; (b) LFS SF- reference configuration; (c) LFS SF+ reference configuration; (d) HFS SF+ reference configuration; (e) HFS SF- reference configuration. A consistent color scheme is adopted that blue for the LFS SF– configuration, orange for LFS SF+, green for HFS SF+, and purple for HFS SF– in this whole paper.*

*Table 1 Parameters of the SN reference and the four types of SF reference equilibria. For the SN equilibrium, the values of dxx, σ and θ are not defined, as there is only one X-point. In the SF+ equilibria, $dxx$ is also not defined because the secondary X-point lies in the private flux region (PFR) and cannot be directly mapped to the outer midplane (OMP) along the magnetic field line.*

| Equilibrium | SN Reference | LFS SF- Reference | LFS SF+ Reference | HFS SF+ Reference | HFS SF- Reference |
|---|---|---|---|---|---|
| Major Radius $R$ | 4.96 m | 4.98 m | 4.97 m | 4.97 m | 4.98 m |
| Minor Radius $a$ | 1.69 m | 1.72 m | 1.74 m | 1.74 m | 1.71 m |
| Plasma Current $I_p$ | 10 MA | 10 MA | 10 MA | 10 MA | 10 MA |
| Magnetic Field on axis $B_T$ | 6.05 T | 5.02 T | 5.03 T | 5.03 T | 5.03 T |
| $q_{95}$ | 6.47 | 5.56 | 5.42 | 5.15 | 4.79 |
| $dxx$ | / | 2.5 mm | / | / | 2.5 mm |
| $\sigma$ | / | 0.50 | 0.27 | 0.43 | 0.64 |
| $\theta$ | / | 39.5° | 51.9° | 88.9° | 117.6° |

## 2.2. Series of snowflake equilibria with varying magnetic geometries

To investigate magnetic geometry effect, the positions of the secondary X-point are systematically scanned across the four types of snowflake configurations. In the SF− configurations, the secondary X-point is located in the SOL region, and experiments [11] have shown that the distance *dxx* can affect in-out power sharing and power splitting at the secondary X-point. Therefore, the secondary X-point in the LFS SF− and HFS SF− configurations is moved horizontally along the R direction, with *dxx* ranging from 1.1 mm to 7.3 mm and from 2.5 mm to 15.3 mm, respectively.

For the SF+ configurations, we focus on the effect of $\sigma$ in this study following the approach in [31]. In this situation, the secondary X-point is moved along the straight line defined by itself and the primary X-point, so that $\theta$ remains constant. In the LFS SF+ and HFS SF+ configurations, $\sigma$ ranges from 0.20 to 0.51 and from 0.29 to 0.64, respectively. These equilibria are employed in the SOLPS-ITER simulations in Section 5 and more details about the equilibria are summarized in Figure 13.

It should be noted that, according to the definition of a snowflake divertor [6][12], the snowflake configuration is realized only when two X-points are in very close proximity. In this study, however, the term is employed in a more relaxed manner with increasing σ, serving as a convenient label.

## 3. Modelling setup

The orthogonal high-quality plasma computational grids, together with neutral triangular meshes for the reference equilibria, are shown in Figure 2. The corresponding sizes of the plasma grids for the SN, LFS SF-, LFS SF+, HFS SF+ and HFS SF- reference cases are 120×38, 164×42, 130×34, 148×44 and 160×46 respectively.

In this work, the plasma computational domain is expanded as much as possible in SOL and PFR, so as to extend over serval density, temperature and power decay lengths $\lambda_n$, $\lambda_T$ and $\lambda_q$, respectively. As an example, the SOL mesh width at the OMP is ~100 mm, whereas the $\lambda_n$ and $\lambda_n$ are below 30mm in all cases and, in some situations, only a few millimeters. The large plasma domain can eliminate the effect introduced by the finite grid size [32]. The core width, which is from the core boudanry to the separatrix measuring at the OMP, is also fixed approximately 100 mm in all cases.

Compared to SN divertor which has three regions(Core, SOL and PFR), there are six regions for the SF divertors. The following nomination rules are adopted in this work and can be illustrated by Figure 3.

1. For SF− divertors as Figure 3(b)(e), the secondary X-point is located in the scrape-off layer (SOL) region of the primary X-point, dividing the SOL into two layers. The layer adjacent to the core plasma is referred to as the SOL1 region, while the outer layer is referred to as the SOL2 region. The PFR is divided into three parts. Starting from the LFS, these regions are named PFR1, PFR2, and PFR3 in a clockwise direction.

2. For SF+ divertors as Figure3 (c)(d), the secondary X-point is located in the PFR of the primary X-point, dividing it into four distinct regions. The region between the primary and secondary X-points is defined as the PFR1 region. Among the remaining three regions, starting from the LFS, they are named PFR2, PFR3, and PFR4 in a clockwise direction.

Compared to SN divertors which has two divertor targets , SF divertors have four targets. In this study, the following nomination rules are adopted and can be illustrated by Figure 3:

1. For LFS SF- divertors as Figure 3 (b), there are one inner target named as IT1 and three outer targets named as OT1, OT2, and OT3. OT1 and OT2 are the main targets and OT1 is connetd with SOL1.
2. For LFS SF+ and HFS SF+ divertors as Figure 3 (c)(d): there are two inner targets and two outer targets. The targets connected with SOL region are named IT1 and OT1 The remaining targets are named IT2 and OT2.
3. For HFS SF- divertors as Figure 3 (e), there are three inner targets named as IT1, IT2 and IT3 and one outer target named as OT1. IT1 and IT2 are the main targets and IT1 is connetd with SOL1.

As a starting point, we consider a simplified divertor target geometry in which the target plates are flat, analogous to those used in TCV[10]. The flat divertor targets are used for the simulations in section 4 and section 5. In Section 6, the ITER-like divertor [33] is employed to study the effect of geomerty, further details are provided in that section.

In this work, only pure deuterium plasma without currents or drifts are considered. The input power at the core boundary is assumed to be 20MW, equally shared between electrons and ions. At the core boundary, a fixed density condition is applied instead of gas puffing fueling, as it offers better numerical stability and faster convergence speed. Convergence criteria are described in [34], and a good particle balance that the pumping rates equals the core fueling rate is checked for all the simulations. Due to the large device size and complexity of magnetic geometry, the convergence time can be several months, even when the neutral gas transport calculation is parallelized. Future simulations will include gas puffing for deuterium fueling and impurity seeding. The pumping is mimicked by surfaces with albedo 0.99 which are marked as red segments in Figure 2. Perpendicular transport coefficient $D_\perp = 0.1\ m^2 s^{-1}$ , $\chi_{\perp,ion} = \chi_{\perp,electron} = 0.3\ m^2 s^{-1}$ are used in the plasma domain of all simulations, resulting in $\lambda_q \sim 6mm$. The physical and numerical models from previous AUG simulation study [35][36], e.g. flux limiter, atomic reactions, surface property etc., are applied in this work.

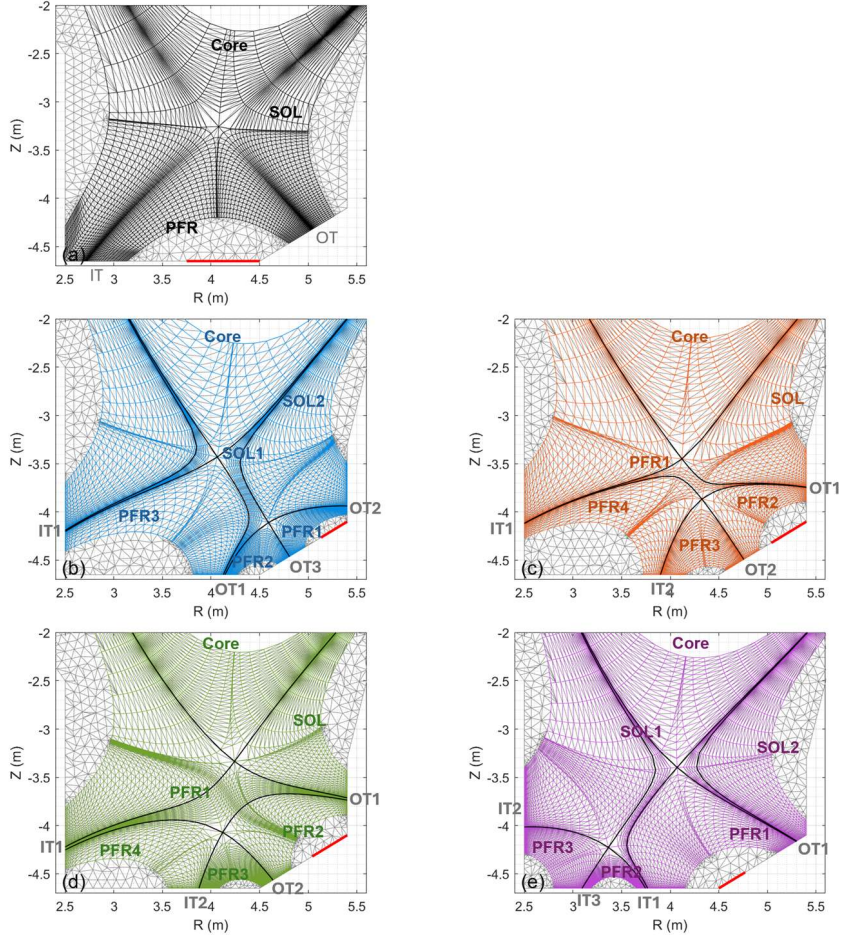

*Figure 2 Computational meshes for (a) the SN reference case and the four types of Snowflake divertors: (b) LFS SF- reference case; (c) LFS SF+ reference case; (d) HFS SF+ reference case and (e) HFS SF-reference case. The red elements represent the pumping surface.*

## 4. Simulation results of reference cases

In this section, a comparative analysis across the simulation results of the SN and four SF reference cases is carried out. The secondary X-point appearing in the SF divertors makes direct modelling of the divertor detachment particularly challenging, as it involves both complex detachment physics and complex magnetic geometry. We start from low upstream density ($n_{core}$=3.0×10$^{19}$ m$^{-3}$, fixed as core boundary conditions for all reference cases) and perform a density scan designed to transition from the low-recycling regime to the high-recycling regime and/or the detachment. It should be noted that the low-density condition $n_{core}$=3.0×10$^{19}$ m$^{-3}$ may not occur in an actual ITER discharge; it is considered here to easily compare the simulation results with the theoretical Two-Point Model (TPM) [38].

### 4.1. Low upstream density condition

The OMP profiles shown in Figure 3 include electron density $n_e$, electron temperature $T_e$ and ion temperature $T_i$. The inner and outer target profiles, including $n_e$, $T_e$, parallel heat load $q_{\parallel}$ and target heat load $q_{surf}$ are shown in Figure 4 and Figure 5. The $q_{surf}$ corresponds to equation (5), and $q_{\parallel}$ is derived from $q_{surf}$ by projecting it onto the parallel direction. All the target profiles presented in this work are mapped to the OMP according to the poloidal magnetic function flux. The 2D distributions of $n_e$, $T_e$ and particle source $S_{D+}$ (ionizations) and particle loss $L_{D+}$ (recombinations) are shown in Figure 6. For all 2D distribution plots, due to the limited range of the color bar, some low-magnitude values in the plots may not be displayed. In Figure 3, for LFS SF+, HFS SF+ and HFS SF- reference cases, the OMP profiles are similar and the separatrix density $n_{e,sep} \sim 5 \times 10^{18}$ m$^{-3}$ and the separatrix temperature $T_{e,sep} \sim 1000$ eV. For the SN and LFS SF- reference cases, there are similar $n_{e,sep} \sim 1.6 \times 10^{19}$ m$^{-3}$ and $T_{e,sep} \sim 200$ eV, while in the far SOL region where $R-R_{sep}>10$mm, SN reference case has much higher $n_e$ than LFS SF- reference case.

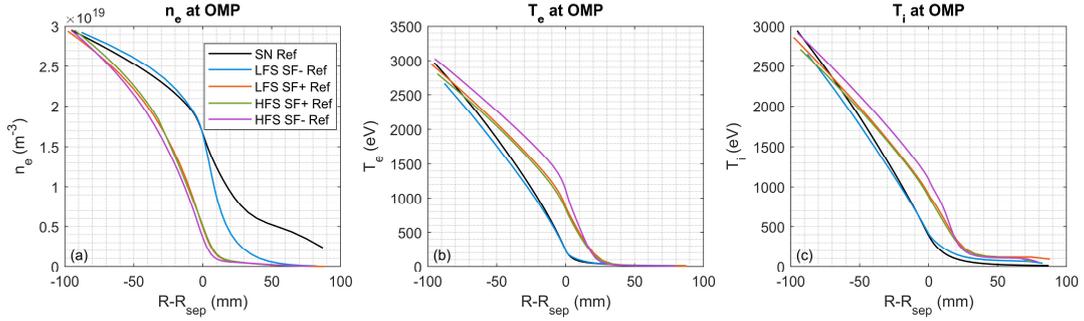

Figure 3 OMP profiles of the SN and the four SF reference cases at low upstream density $n_{core}=3.0\times10^{19}$ m$^{-3}$. (a) electron density $n_e$; (b) electron temperature $T_e$ and (c) ion temperature $T_i$.

The discrepancies of OMP $n_e$ profiles are due to the synergistic effect from magnetic geometry and divertor target geometry. For the SN reference case, the inner target (IT) has a horizontal–vertical shape as 'corner slot' [37], which helps to baffle neutrals in the HFS SOL region. The poloidal tilting angle $\alpha$ at the outer target (OT) is approximately 90° such that neutrals are recycled directly back to the LFS SOL. These features result in a strong ionization source $S_{D+}$ in the LFS and HFS upstream as in Figure 6 (c) and (d), even recombination source (Particle loss source) $L_{D+}$ appears in the HFS SOL. Correspondingly, there is high upstream density, the IT is partially detached, and the OT is in the high-recycling regime as in Figure 4 (a) and (b) and Figure 5 (a) and (b).

For the LFS SF− reference case, it is interesting to observe that a recombination zone formed in the PFR3 region and closed to the LFS SOL1 region, as can be seen in Figure 6 (h). On the one hand, due to the presence of the secondary X-point, the distance between HFS SOL and LFS SOL1 is close. For example, in the SN reference case, the distance between the inner and outer target strike points is approximately 2 m. In contrast, for the LFS SF− configuration, the distance between the strike points on IT1 and OT1 is reduced to about 1.2 m. As a result, recycling neutrals from the vertical inner target can more easily enter PFR3 and even LFS SOL1. On the other hand, the secondary X-point also splits the power at the outer divertor entrance, so that only 3.8 MW power enters LFS SOL1 channel and 8.5 MW power enters LFS SOL2 channel. Based on these two aspects, there is the formation of the recombination zone in PFR3 and the total recombination rate is $1.98\times10^{23}$ s$^{-1}$. According to the target profiles in Figure 5 (e) and (f), the OT1 is in the high recycling regime and the recycling neutrals from OT1 can go to SOL2, PFR1 and PFR2 that result in strong $S_{D+}$ in Figure 6 (g). However, due to the vertical geometry of OT2 and the high electron temperature in PFR1, the recycling neutrals are primarily ionized in front of the OT2 divertor target and within the PFR1 region, preventing them from propagating further upstream. The LFS-SF- ionization source in the upstream far-SOL from Figure 6(g) is smaller than that in the SN case from Figure 6(c), which explains the higher density observed in SN from R=30mm to 100mm in Figure 3(a).

In the LFS SF+ and HFS SF+ cases, due to the magnetic geometry and vertical targets, ionization primarily occurs in the PFRs near the secondary X-point rather than in the SOL which are shown in Figure 6(k) and (o). This prevents recycling neutrals from effectively entering SOL, hindering the transition to a high-recycling regime. Thus, the IT1 and OT1 are in the low-recycling regime which are inferred from target $T_e$ profiles Figure 5 (j)(n).

For the HFS SF− reference case, the poloidal tilting angle α of OT1 is also ~90°. However, recycling neutrals from IT1 and IT2 cannot go to the LFS SOL region. Because of the high $T_e$ in the HFS SOL2, recycled neutrals are mostly ionized in front of IT2 and around the secondary X-point as Figure 6(s). Therefore, they cannot enhance the recycling at OT1 as effectively as in the LFS SF− case. As a result, the targets are in the low-recycling regime as shown in Figure 4(r) and Figure 5 (r).

When the targets are in the low recycling regime, the $T_e$ at the target matches well with the upstream temperature that $T_{up} = T_{target}$, as observed for LFS SF+ IT1 and OT1, HFS SF+ IT1 and OT1, as well as HFS SF- IT1, IT2, and OT1 in Figure 4 and Figure 5. However, the corresponding target $n_e$ profiles, deviate from what predicted by the TPM: $n_{target} = \frac{1}{2} n_{up}$ in the low-recycling regime. This deviation arises because the TPM assumes recycled neutrals are emitted and re-ionized within the same flux tube. In SF divertor with low upstream density, cross-field transport can lead to ion motion into adjacent flux regions (PFRs), where the $T_e$ remains high. As a result, recycled neutrals may be ionized in these neighboring regions rather than within the original flux tube. Under such circumstances, the TPM density relation breaks, leading to cases where the target density is lower than one half of the upstream density. This is also consistent with the 2D distributions of $n_e$ and $S_{D+}$ in Figure 6, showing high values near the secondary X-point in LFS and HFS SF+.

For the heat flux profiles at the inner targets, the advantages of the HFS SF- divertor are evident from Figure 4. The peak parallel heat load $q_{\parallel}$ in the LFS SF+, HFS SF+, and HFS SF- cases is approximately 160 MW/m². However, the peak surface heat flux $q_{surf}$ in the HFS SF+ case is around 6 MW/m², which is about 50% of that in the other types of SF divertor. For the outer targets, the advantages of LFS SF- divertor is obviously that the peak value of $q_{surf}$ ~ 6 MW/m² which is lower than other cases as shown in Figure 5 (h).

Under the low upstream density condition, the particle throughput for LFS SF- is $4.7 \times 10^{20}$ atom/s which is much higher than in the other SF cases ($3.1–7.5 \times 10^{19}$ atom/s). This is consistent with the fact that the LFS SF⁻ divertor operates in the high-recycling regime, whereas the others remain in the low-recycling regime.

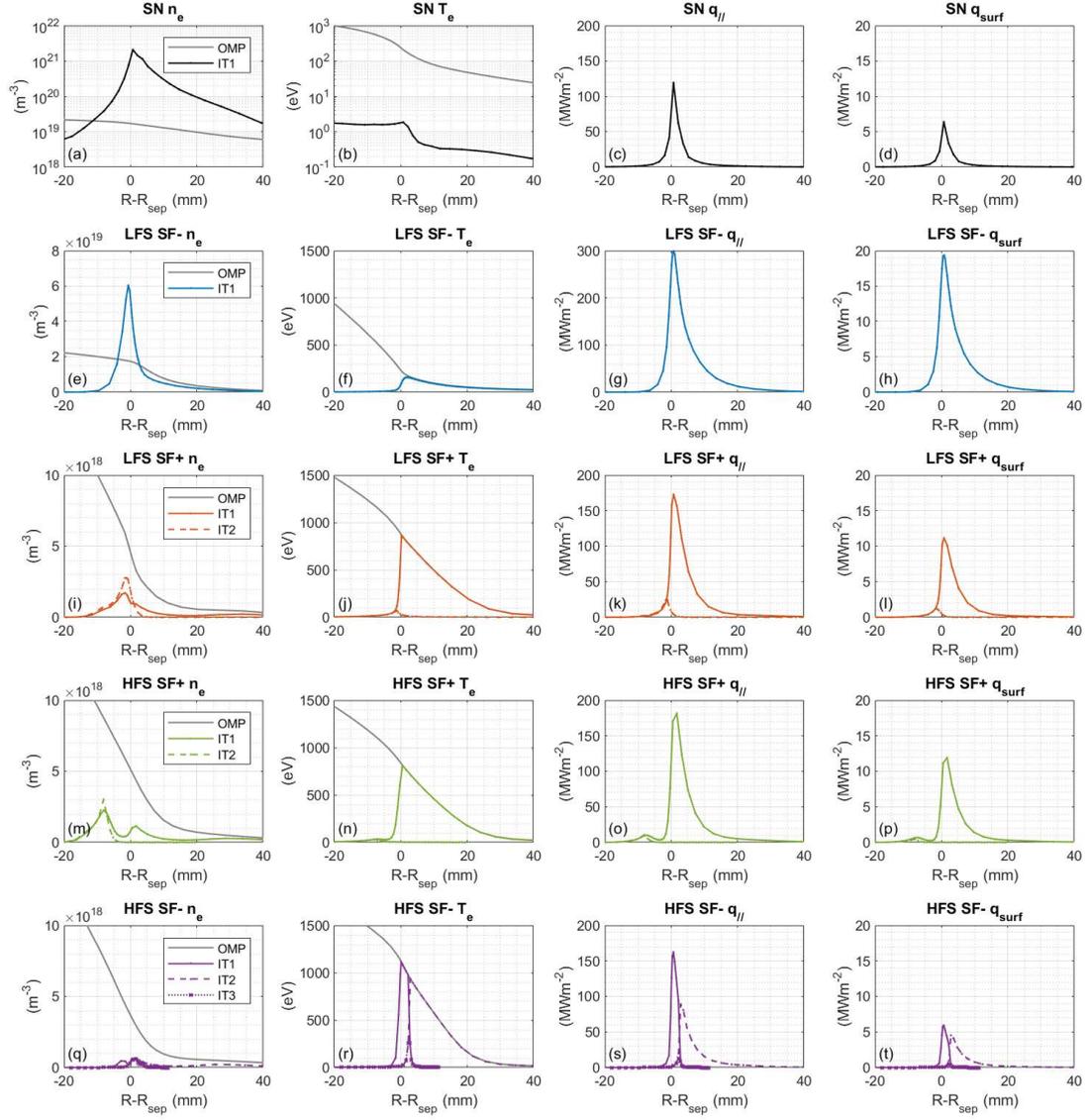

*Figure 4 Inner target profiles of the SN and the four SF reference cases at the low upstream density condition: $n_{core}=3.0\times10^{19}$ m$^{-3}$. From left to right columns are electron density $n_e$, electron temperature $T_e$, parallel heat load $q_{//}$ and target heat load $q_{surf}$. From the top to the bottom rows are the SN, LFS SF-, LFS SF+, HFS SF+, HFS SF- reference cases. For the $n_e$ and $T_e$, the OMP profiles are also presented as a comparison. The values are mapped to the OMP.*

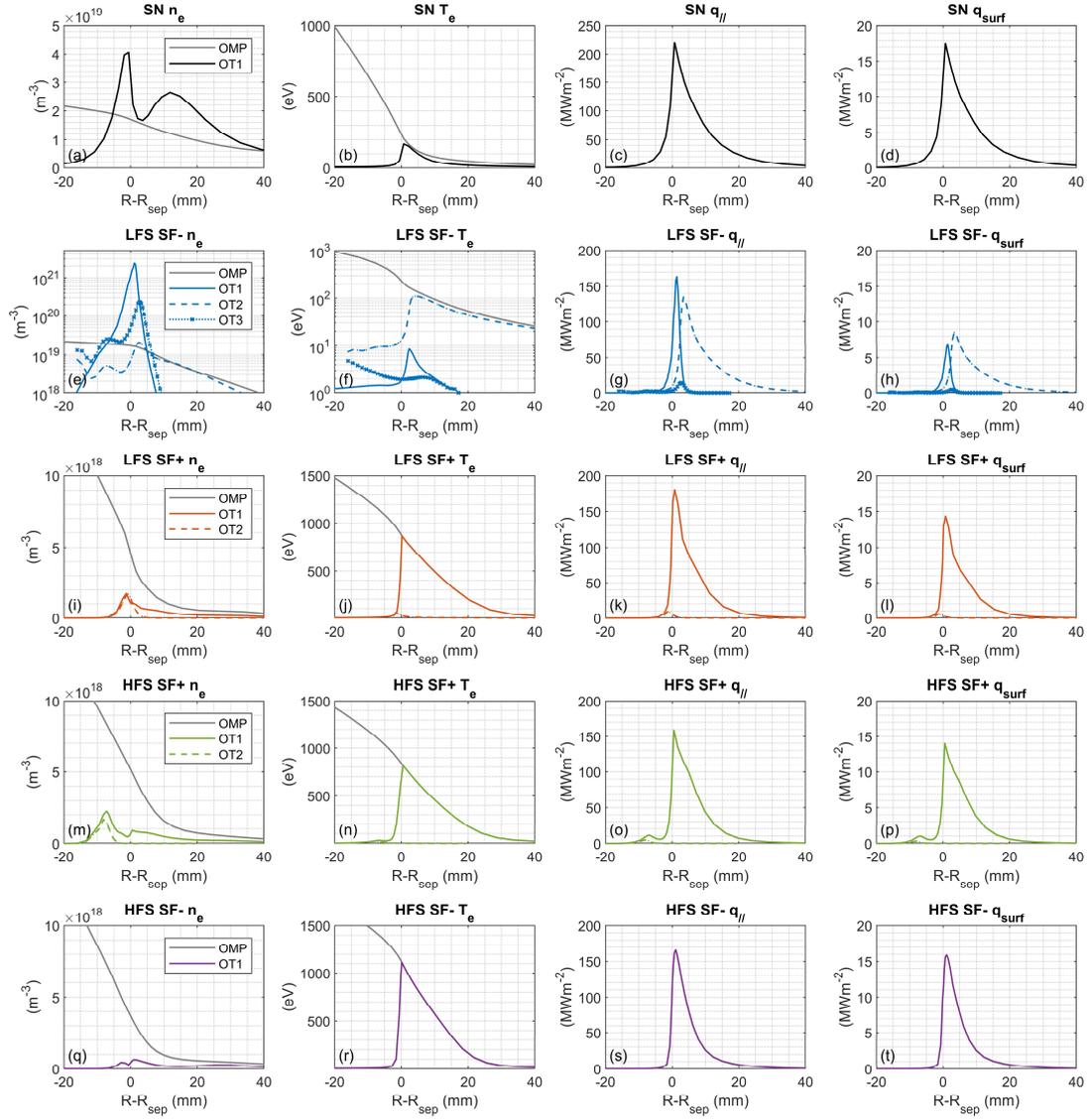

*Figure 5 Outer target profiles of the SN and the four SF reference cases at the low upstream density condition: $n_{core}=3.0\times10^{19}$ m$^{-3}$. From left to right columns are electron density $n_e$, electron temperature $T_e$, parallel heat load $q_{//}$ and target heat load $q_{surf}$. From the top to the bottom rows are the SN, LFS SF-, LFS SF+, HFS SF+, HFS SF- reference cases. For the $n_e$ and $T_e$, the OMP profiles are also presented as a comparison. The values are mapped to the OMP.*

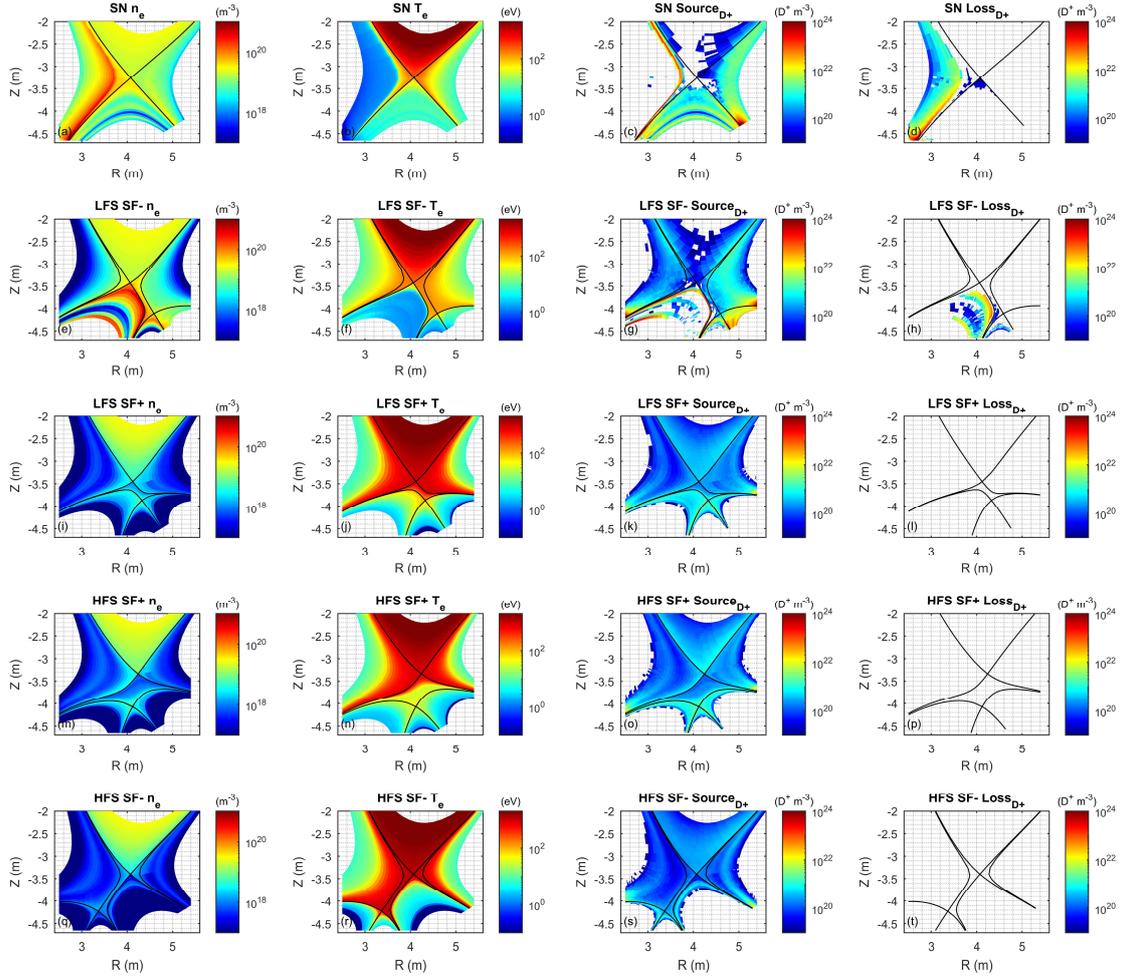

*Figure 6. 2D distribution of plasma parameters of the SN and the four SF reference cases at the low upstream density condition : $n_{core}=3.0\times10^{19}$ m$^{-3}$. From the left to the right columns are electron density $n_e$, electron temperature $T_e$, particle source $S_{D+}$, particle loss $L_{D+}$ due to recombination. From the top to the bottom rows are the SN, LFS SF-, LFS SF+, HFS SF+, HFS SF- reference cases. For LFS SF+, HFS SF+ and HFS SF- cases, the values of the $L_{D+}$ are too small and out of color bars that are not displayed here.*

### 4.2. Upstream density scan

Starting from $n_{core}=3.0\times10^{19}$ m$^{-3}$, upstream density scans are performed for the SN and SF cases, increasing the value up to $5.0\times10^{19}$ m$^{-3}$. The upstream separatrix density $n_{e,sep}$ and temperature $T_{e,sep}$ as a function of $n_{core}$ are shown in Figure 7. As the core density increases, the difference between the cases decreases and cancels in the high core density region that $n_{e,sep} \sim 3.0\times10^{19}$ m$^{-3}$ and $T_{e,sep} \sim 200$eV, respectively, for all cases.

The electron temperature at the strike points and the peak parallel heat load at the targets are summarized in Figure 8. We select them as key parameters to assess the divertor performance. Due to the limited space in the figure, the values of LFS SF- OT3, LFS SF+ IT2 and OT2, HFS SF+ IT2 and OT2 and HFS SF- IT3 are not presented because heat load at these targets can be neglected. The advantages of the LFS SF− divertor become more pronounced as the upstream density increases, as it exhibits the lowest electron temperatures at the outer target. For inner target, the HFS SF- divertor has better performance than others except the SN divertor. This is because the horizontal–vertical shape of SN inner divertor which has been discussed in Section

2.1. Among all SF divertors, the outer targets in LFS SF− and the inner targets in HFS SF− can be more easily detached. It is also found that the secondary X-point affects the power in-out sharing. In fact, Figure 8 shows that at the inner target the peak value of $q_{\parallel}$ in HFS SF+ is lower than the one LFS SF+ and vice versa at the outer target.

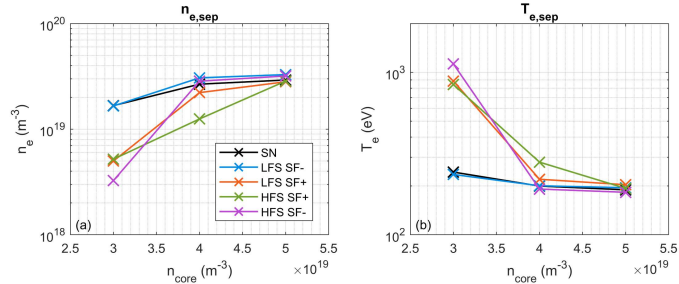

Figure 7 (a) Upstream separatrix density $n_{e,sep}$ and (b) temperature $T_{e,sep}$ as a function of $n_{core}$ for the SN and the four SF reference cases.

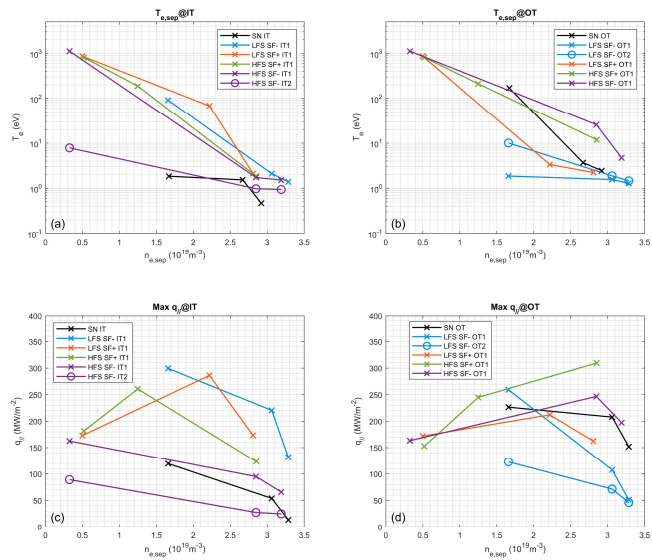

Figure 8 Key plasma parameters as a function of a function of $n_{e,sep}$. The top row is electron temperature at the target strike point and the bottom row is maximum parallel heat load at the target. The left column is for the inner target and the right column is for the outer target.

### 4.3. High upstream density condition

The OMP profiles under the high upstream density condition $n_{core}=5\times10^{19}$ m$^{-3}$ are shown in Figure 9. For $n_{e,sep}$, the SN and four SF reference cases reach similar levels ∼3.0×10$^{19}$ m$^{−3}$, and the discrepancies observed in Figure 4 disappear. This is because, in high upstream density conditions, the divertor targets which are connected to the SOL regions, i.e. IT1, OT1 and OT2, of LFS SF- and IT1 and OT1 of SF+ divertors, enter the high-recycling regime or detached, and a drop in the local temperature near the targets appears. The ionization source shifts from the PFRs, e.g. the PFR2, PFR3 and PFR4 of SF+ divertors, toward the SOL regions as shown in Figure 12 (k)(o). With similar level of ionization source in the SOL, the SN and SF references cases have similar OMP profiles.

The inner and outer targets profiles are shown Figure 10 and Figure 11. For SN reference case, the IT is fully detached, with peak values of $q_\parallel$ ~20 MWm$^{-2}$ and $q_{surf}$ ~1MWm$^{-2}$. This is due to its structure as mentioned in previous discussion. The OT is in the onset of detachment which is inferred from $T_e$ profiles in Figure 11 (b) and the peak values of $q_\parallel$ ~150 MWm$^{-2}$ and $q_{surf}$ ~10 MWm$^{-2}$.

For the LFS SF- reference case, as the upstream density increases, the IT1 target enters from low recycling regime to the onset of detachment. In the region where $R-R_{sep}$ is 0–5 mm, the $T_e$ falls below 5 eV, as shown in Figure 10 (f). Compared to the low upstream density case in section 4.1, more recycling neutrals from IT1 enter the PFR3 region and are subsequently ionized in the region around the interface between SOL1 and PFR3. The recombination zone observed at low upstream density case becomes more intense and expands to the whole SOL3 region and to the SOL1 region, as shown in Figure 12 (h). Both ionization and recombination increase the volumetric dissipation. The OT1, which is magnetic connected with the ionization and recombination zones, is fully detached as shown in Figure 11 (f) that $T_e$ falls below 2eV. The recycling neutrals from OT1 lead to a strong recombination source formed in the PFR1, PFR2 and the region close to OT2, facilitating the detachment of OT2. The total recombination is 5.44×10$^{23}$ s$^{-1}$, 5 times higher than the value in the low upstream density case. The peak values of $q_\parallel$ at OT1 and OT2 are ~50MWm$^{-2}$ and the peak $q_{surf}$ are ~3MWm$^{-2}$ and only ~30% of those in the SN reference case. This reveals a noticeable consequence of the LFS SF− divertor: the presence of the secondary X-point shortens the distance between IT1 and OT1, allowing recycling neutrals from IT1 to more easily enter the PFR3 and LFS SOL1 regions compared to the SN divertor. Benefitting from the enhanced neutral content, the volumetric dissipation increases due to ionization and recombination. As a result, OT1 can achieve full detachment, leading to a significant reduction of the target heat load $q_{surf}$.

For the LFS SF+, the peak value of $q_\parallel$ at the IT1 and OT1 are 170MWm$^{-2}$ and 120 MWm$^{-2}$, respectively and they are in the high recycling regime which are inferred from $T_e$ profiles in Figure 10 (j) and Figure 11 (j). However, for the HFS SF+, the peak value of $q_\parallel$ at the IT1 and OT1 are 50MWm$^{-2}$ and 200 MWm$^{-2}$. The IT1 target is partially detached, with the $T_e$ falling below 2 eV in the region where $R-R_{sep}$ is 0–5 mm as in Figure 10 (n). In contrast, the OT1 target is in a high-recycling regime, with $T_e$ ranging from 10 to 50 eV in the same region as in Figure 11 (n). These observations are thought to be related to the plasma–neutral interaction occurring in the PFR1. In Figure 12 (m), It can be seen that there is a high-density zone between the primary and secondary X-points, which is consistent with a previous EMC3-EIRENE study [31]. In the PFR1, strong ionization and recombination zones exist as in Figure 12 (o) and (p). The zones are close to the primary X-point. This feature is believed to be related to the long connection length and flux expansion in the PFR1 region, which result from the presence of the secondary X-point. The presence of strong ionization and recombination source near the primary X-point may offer potential advantages to the formation of the X-point radiator [39][40]. This aspect will be investigated in more detail, especially the impurity seeding simulation in future work.

For the HFS SF− reference case, the peak parallel heat load $q_\parallel$ is approximately 60 MW/m² at IT1 and 20 MW/m² at IT2 in Figure 10 (s). The underlying processes are like those in the LFS SF− reference case; the main difference being that the recombination source forms on the HFS side instead of the LFS side. The heat load $q_{surf}$ is comparable to that at the IT of SN reference case and are significantly lower than those in the other three SF cases. These results indicate that the HFS SF- divertor is favorable for power exhaust at the inner target.

Under the high upstream density condition, the particle throughput for the SF cases is of the same order of magnitude, ranging from $1.9\times10^{21}$ to $3.5\times10^{21}$ atom/s. Compared with the low upstream density cases, these values are higher by 1–2 orders of magnitude, which is consistent with the fact that at high upstream density the targets are in the high-recycling regime or detached. In the future, deuterium gas puffing will be considered instead of prescribing a fixed core density, which will ensure that the particle throughput is the same across different SF divertors. And the divertor performance will be evaluated under same particle throughput conditions.

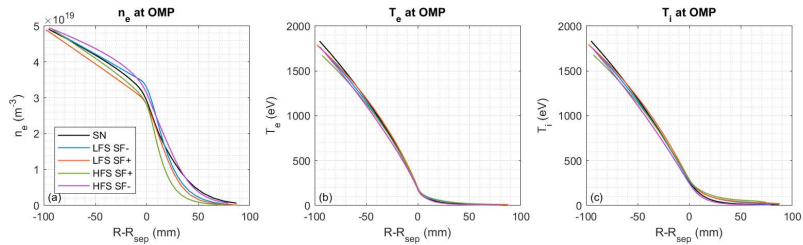

Figure 9 OMP profiles of the SN and the four SF reference cases at the low upstream density condition. (a) the electron density $n_e$; (b) electron temperature $T_e$ and (c) ion temperature $T_i$.

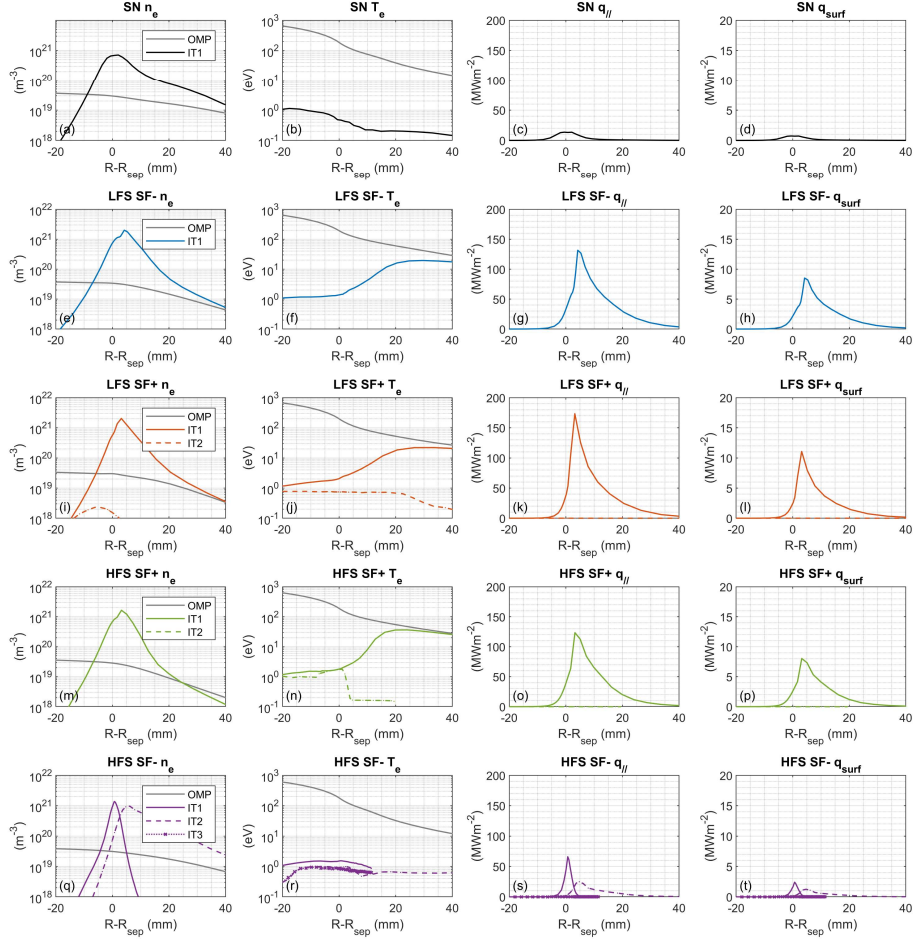

*Figure 10 Inner target profiles of the SN and the four SF reference cases at the low upstream density condition: $n_{core}=5.0\times10^{19}$ $m^{-3}$. From left to right columns are electron density $n_e$, electron temperature $T_e$, parallel heat load $q_{//}$ and target heat load $q_{surf}$. From the top to the bottom rows are the SN, LFS SF-, LFS SF+, HFS SF+, HFS SF- reference cases. For the $n_e$ and $T_e$, the OMP profiles are also presented as a comparison. The values are mapped to the OMP.*

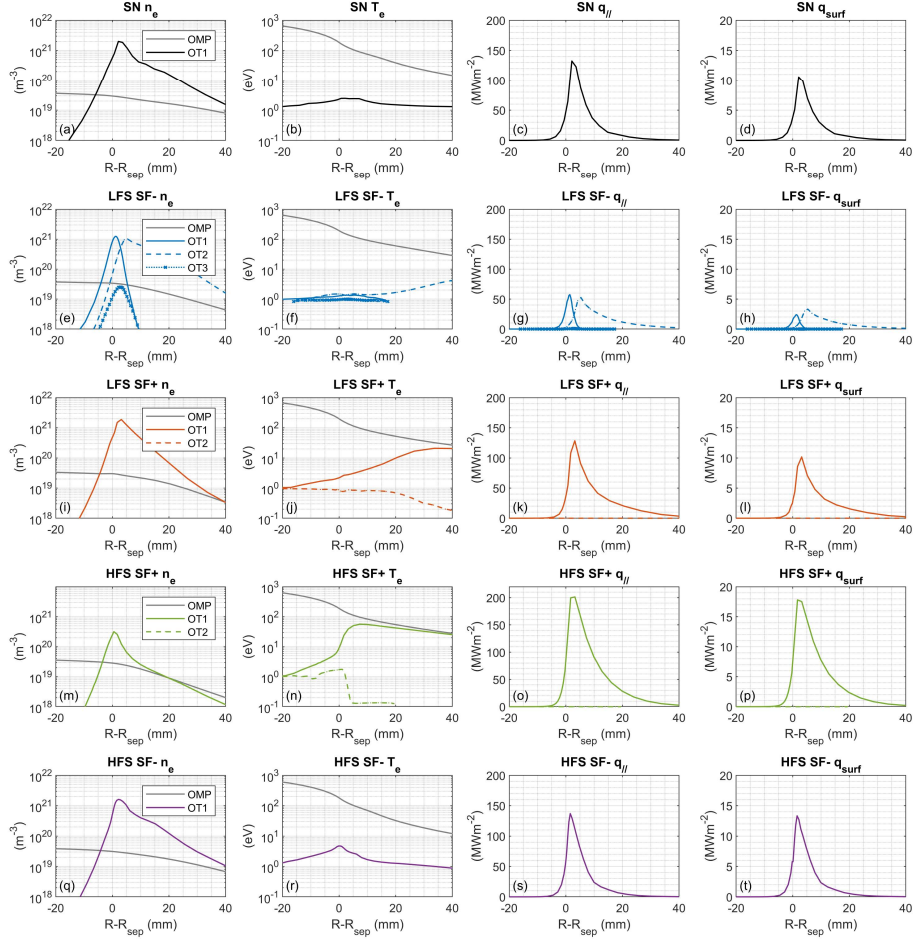

Figure 11 Outer target profiles of the SN and the four SF reference cases at the low upstream density condition that $n_{core}$=5.0E19 $m^{-3}$. From left to right columns are electron density $n_e$, electron temperature $T_e$, parallel heat load $q_{//}$ and target heat load $q_{surf}$. From the top to the bottom rows are the SN, LFS SF-, LFS SF+, HFS SF+, HFS SF- reference cases. For the $n_e$ and $T_e$, the OMP profiles are also presented as a comparison. The values are mapping to the OMP.

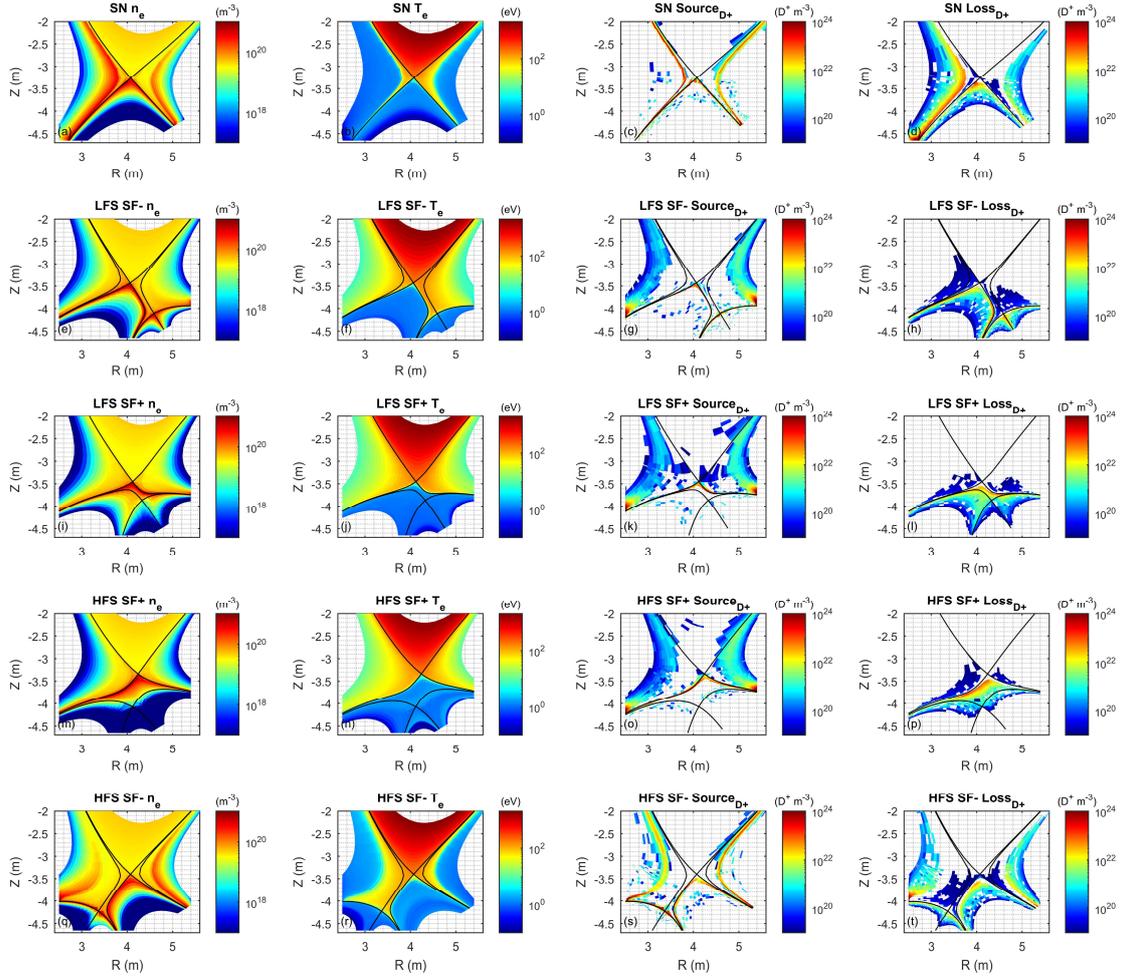

*Figure 12 2D distribution of plasma parameters of the SN and the four SF reference cases at the high upstream density condition. From the left to the right columns are electron density, electron temperature, particle source $S_{D+}$, particle loss $L_{D+}$ due to recombination. From the top to the bottom rows are the SN, LFS SF-, LFS SF+, HFS SF+, HFS SF- reference cases.*

## 5. Effect of magnetic geometry

In this section, the effect of magnetic geometry on plasma behavior and power exhaust performance in each type of SF divertors are studied by scanning the secondary X-point position as mentioned in section 2.2. The computational meshes which correspond to the equilibria in section 2.2, together with dxx, σ and θ values, are summarized in Figure 13. For LFS and HFS SF- divertors the X-points distance *dxx* is scanned and for LFS and HFS SF+ divertors the normalized distance *σ* is scanned. As *dxx* and *σ* changes, the sizes of structured plasma mesh in the SOL1 region for SF− divertors and in the PFR1 region for SF+ divertors are adjusted to maintain comparable spatial resolution and exclude the numeric discretization effect. All the cases in this section use the high upstream density condition: $n_{core}$=5×10$^{19}$ m$^{-3}$. The LFS SF- Case2, LFS SF+ Case2, HFS SF+ Case2 and HFS SF- Case1 are the reference cases which were discussed in section 4.3. In high upstream density condition, for each type of SF divertors, they have similar OMP density and temperature profiles, e.g. from LFS SF- Case1 to LFS SF- case5. Due to the limited space of this paper, the OMP profiles are not presented here.

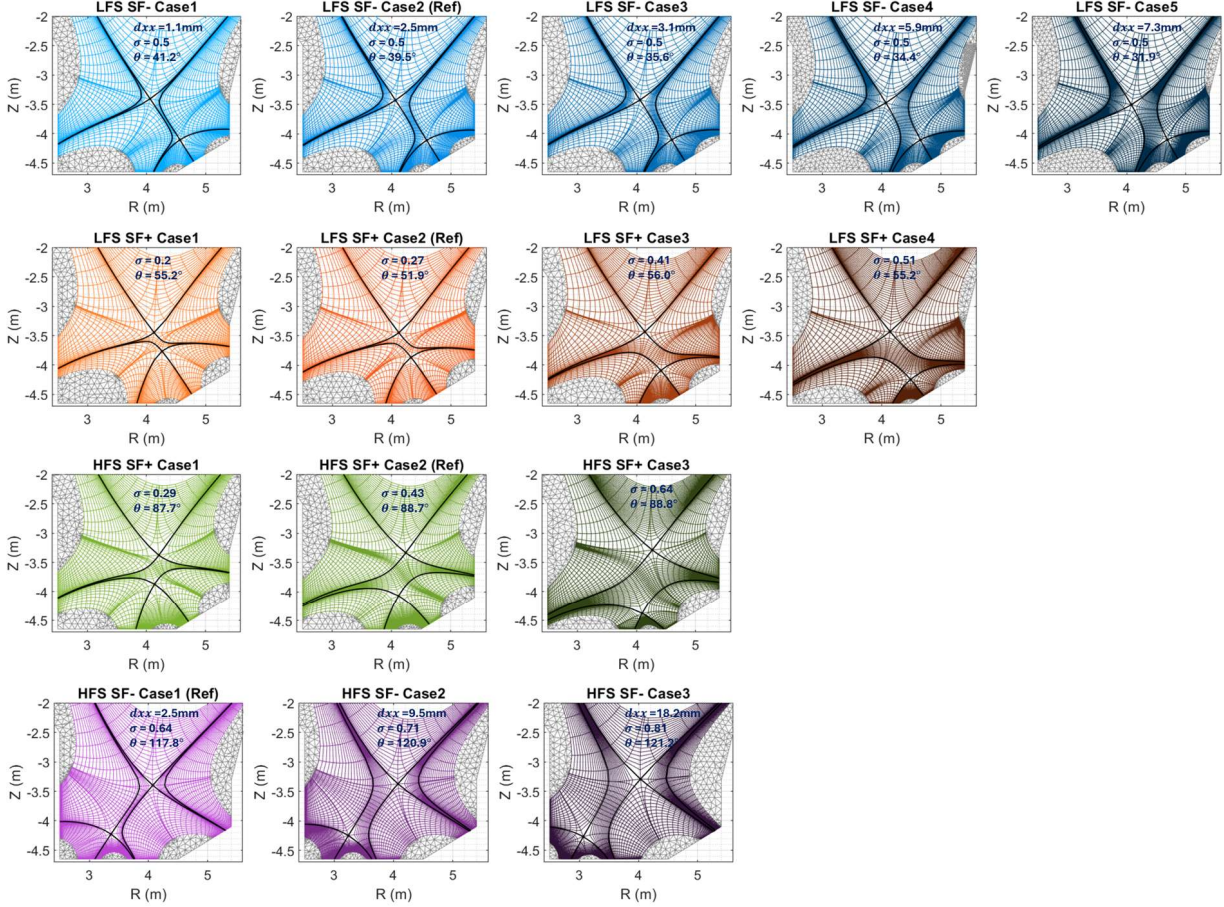

*Figure 13 Computational meshes with dxx, σ and θ values corresponding to the scan of secondary X-point positions are constructed based on the equilibria in Section 2.2. From top to bottom, the rows represent the LFS SF−, LFS SF+, HFS SF+, and HFS SF− cases. For the SF− cases, the X-points distance dxx ss is scanned, while for the SF+ cases, the normalized distance σ is scanned. The detailed magnetic geometry information is provided in Figure 2. As dxx and σ changes, the grid sizes in the SOL1 region for SF− cases and in the PFR1 region for SF+ cases are adjusted to maintain the same level of a spatial resolution.*

## 5.1. *dxx* scan in LFS and HFS SF- divertors

The simulation results on power sharing and splitting are shown in Figure 14. $P_{in}$ is the power at the inner divertor entrance and $P_{out}$ is the power at the outer divertor entrance. For the LFS SF− divertor, the presence of the secondary X-point causes the $P_{out}$ to split into two components, which are transported along SOL1 and SOL2, referred to $P_{out,SOL1}$ and $P_{out,SOL2}$, respectively. The power splitting in HFS SF- is similar but happens at the inner divertor entrance, which refers to $P_{in,SOL1}$ and $P_{in,SOL2}$.

For LFS SF- and HFS SF- divertors, as the *dxx* changes, the in-out power sharing ratio remains approximately constant at ~38% and ~32%, respectively, as shown in Figure 14 (a). The constant ratio is not consistent with TCV experimental study [11]. It should be noted that the present simulations do not include impurities or drifts and may affect in-out power asymmetries. These effects on power-sharing will be considered in future work.

For the power splitting, as the *dxx* increases in SF- divertors, more power should transport within SOL1 region. This trend, which has been observed in [11] are reproduced both for the LFS and HFS SF- divertors as in Figure 14 (b).

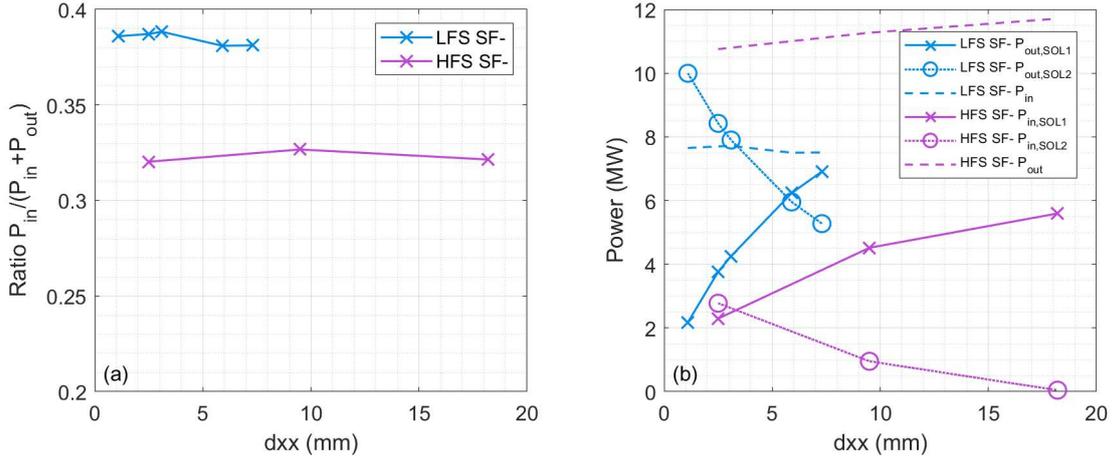

*Figure 14 (a) Power sharing ratio $P_{in}/(P_{in}+P_{out})$ as a function of the distance of the two X-points dxx in LFS and HFS SF- cases; (b) Power entering the LFS divertor region from SOL1 entrance $P_{out,SOL1}$ and from SOL2 entrance $P_{out,SOL1}$ as a function of dxx in the LFS SF− cases, and power entering the HFS divertor region from SOL1 entrance $P_{in,SOL1}$ and from SOL2 entrance $P_{in,SOL2}$ as a function of dxx for the HFS SF− cases.*

The target profiles for the *dxx* scan of the LFS and HFS SF− cases are shown in Figure 15 and Figure 16. For LFS SF- cases, as the dxx increases, even more power goes to SOL1 region, the OT1 and OT2 detach in all cases which can be inferred from the $T_e$ profiles as in Figure 15 (f) and (g). This also indicates that even with increased power entering the SOL1 channel, the volumetric power dissipation due to ionization and recombination, as discussed in Section 4.3, is still sufficient to trigger the detachment of OT1. There is a trade-off between $q_{surf, IT1}$, $q_{surf, OT1}$ and $q_{surf, OT2}$. As the dxx increases, the peak value of $q_{surf, IT1}$ decreases from 10MWm$^{-2}$ to 5MWm$^{-2}$, while at OT1 the peak value of $q_{surf, OT1}$ increases from 1MWm$^{-2}$ to 3MWm$^{-2}$ and at OT2 the peak value of $q_{surf, OT2}$ decreases from 4MWm$^{-2}$ to 1.2 MWm$^{-2}$. This suggests a potential optimization strategy for the LFS SF− divertor: by accurately controlling the two X-points distance *dxx*, it may be possible to achieve a balanced power distribution among all three targets.

For the HFS SF- cases, all the IT1 and IT2 are detached, with the peak $T_e$ below 3eV as in Figure 16 (e) and (f). For case2 and case3, *dxx* (9.5 and 18.1 mm, respectively) is already larger than $\lambda_q$. Consequently, most of the power flows along the SOL1 channel, and there is no obvious difference of $q_{\parallel}$ at IT1. The differences in $q_{surf}$ at OT1 are attributed to the target geometry: in Cases 1 and 2, the far SOL terminates at vertical plates, whereas in Case 3, it ends on a sloped one.

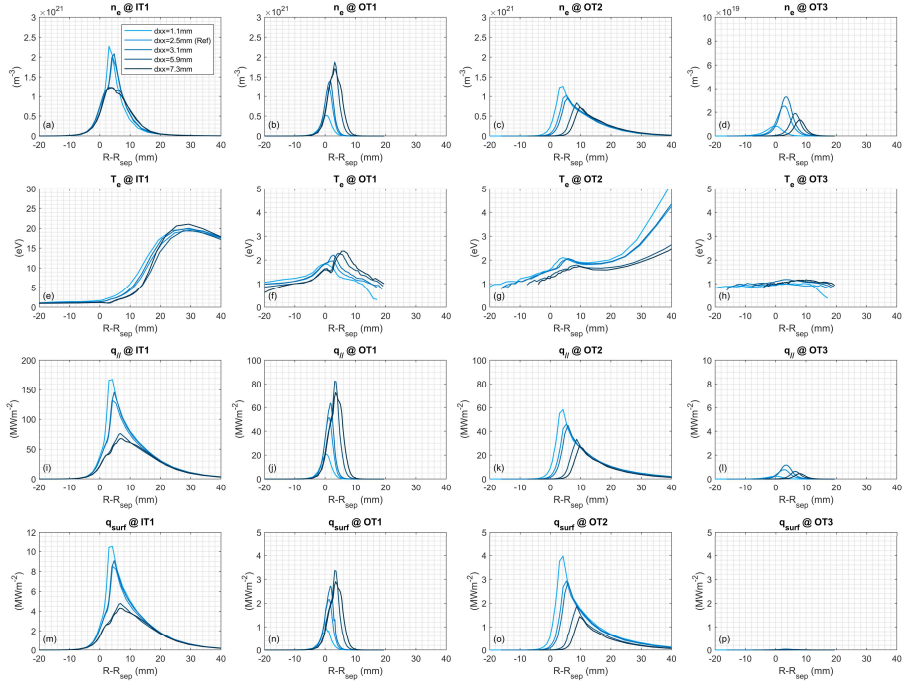

*Figure 15 Target profiles of LFS SF− cases in the scan of dxx. From the top to bottom rows are electron density $n_e$, electron temperature $T_e$, parallel heat load $q_\parallel$, target hear load $q_{surf}$. From the left to the right columns are the profiles at IT1, OT1, OT2 and OT3. The values are mapped to the OMP. The values at the OT3 are negligible.*

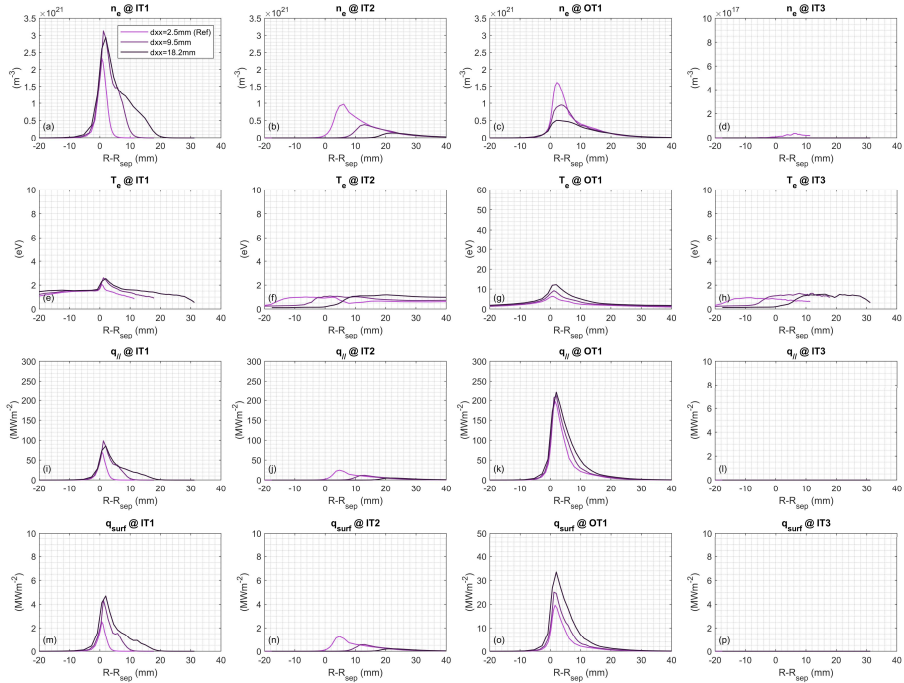

*Figure 16 Target profiles of HFS SF− cases in the scan of dxx. From the top to bottom rows are electron density $n_e$, electron temperature $T_e$, parallel heat flux $q_\parallel$, perpendicular hear flux $q_{surf}$. From the left to the right columns are the profiles at IT1, IT2, OT1 and IT3. The values are mapped to the OMP. The values at the IT3 are negligible.*

## 5.2. σ scan in LFS and HFS SF+ divertors

For SF+ divertors, the secondary X-point is in the PFR and doesn't split the SOL. The normalized distance σ is scanned with unchanged θ value, as mentioned in section 2.2. From the σ scan, the in-out power sharing ratios of LFS SF+ and HFS SF+ are approximately ~33-34% and ~29-

30%, respectively. Such a narrow range indicates that even with a large value of σ, the in–out power sharing changes only slightly. However, when comparing the power at the entrance between the HFS SF+ and LFS SF+ divertors, it is evident that the location of the secondary X-point affects the values of power flux. For example, the $P_{in}$ in LFS SF+ and HFS SF+ are 13MW and 14MW, respectively. We speculate that this might be related to θ, but this guess needs further investigation.

The target profiles are shown in Figure 18. For the LFS SF+ cases, the difference of $q_\parallel$ and $q_{surf}$ are within 20%. The IT1 and OT1 are in the high-recycling regime. For the HFS SF+, the peak value of $q_\parallel$ at IT1 and OT1 are almost the same. The IT1 is partially detached and the OT1 is in the high-recycling regime according to the $T_e$ profiles in Figure 18(c) and (d). No significant improvement in power exhaust performance is observed for SF+ divertors in the pure deuterium simulations.

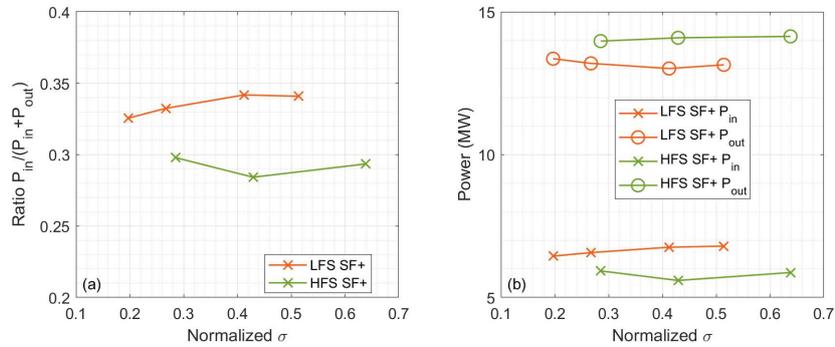

*Figure 17 (a) Power sharing ratio $P_{in}/(P_{in}+P_{out})$ as a function of normalized distance σ in LFS and HFS SF+ cases; (b) Power fluxes at the divertor entrance for LFS SF+ cases and for HFS SF+ cases.*

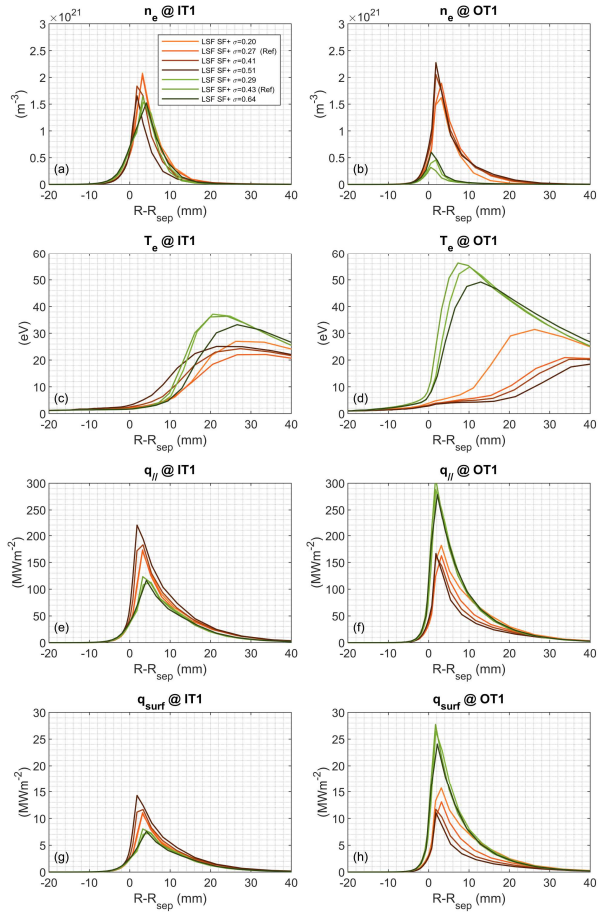

*Figure 18 Target profiles of LFS SF+ and HFS SF+ cases in the scan of σ. From the top to bottom rows are electron density $n_e$, electron temperature $T_e$, parallel heat load $q_\parallel$, target hear load $q_{surf}$. The left column is the profiles at IT1 and the right column is the profiles at OT1. The values at IT2 and OT2 are neglectable and not shown here.*

## 6. Effect of target geometry

The geometry of the target plate plays a critical role in shaping plasma profiles because it affects the neutral recycling dynamics. A closed divertor [41][42] can reflect neutrals in the SOL region; this allows detachment also at low upstream density. The highly inclined target plates also decrease the poloidal tilting angle $α$, which could result in a lower $q_{surf}$. In this section, the SF reference equilibria introduced in Section 2.1 with ITER-like (IL) divertor structure are simulated to investigate the impact of target geometry, by comparing them with reference cases which employ flat target plates. We selected a divertor shape similar to ITER F57 [25] without the dome structure.

The computational meshes with the IL divertor structure are shown in Figure 19. The sizes of the plasma computational grids are the same as the corresponding reference cases in section 4. The IL target shapes in the four SF configurations are the same but have been horizontally and/or vertically shifted to place the strike points at the same position as those in the reference cases in Section 4, thereby minimizing differences in connection length near the separatrix. In the HFS SF- IL case, the outer separatrix lines do not terminate at the outer IL divertor structure. Instead, the outer IL divertor structure acts as a neutral baffle, so that the LFS SOL plasma region remains unaffected. Both the low and high upstream densities are considered by prescribing $n_{core}$ =3×10$^{19}$ m$^{-3}$ and 5×10$^{19}$ m$^{-3}$ and named as IL Low density case and IL High density case respectively, in order to have comparisons with the reference cases from section 4.

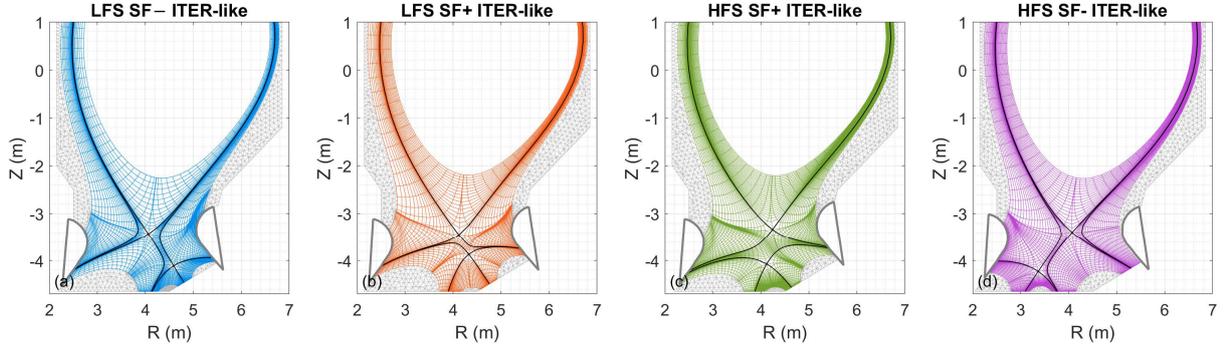

*Figure 19 Computational meshes for (a) LFS SF−, (b) LFS SF+, (c) HFS SF+, and (d) HFS SF− configurations with the ITER-like (IL) divertor structure. The IL divertor shape is indicated by the gray lines. Compared to the reference cases, the strike point positions remain unchanged. In the case of HFS SF−, OT1 is not affected by the IL shape since its strike point does not lie on the IL plates.*

The OMP profiles for the four types of SF IL cases are shown in Figure 20, with the corresponding reference cases included for comparison. It can be found that in the low upstream density condition, the IL target geometry results in an increase of upstream density which is consistent with experiments [42]. However, when the $n_{core}$ = 5×10$^{19}$ m$^{-3}$, with IL geometry, the density decreases for the LFS SF-, LFS and HFS SF+ cases. This is because the power in the SOL region near the target is insufficient to ionize the recycling neutrals. Due to the inclined IL target plate, the recycling neutrals are baffled and tend to flow directly toward the PFR region which is close to the pumping surface. The recycling neutrals are pumped instead of ionization in the SOL region which affects the upstream density. This is confirmed by the HFS SF− IL High density case, in which the outer IL target plate does not redirect recycling neutrals into the PFR region but rather traps them in the SOL region. Thus, HFS SF- IL case exhibits OMP profiles like those of the HFS SF- reference case.

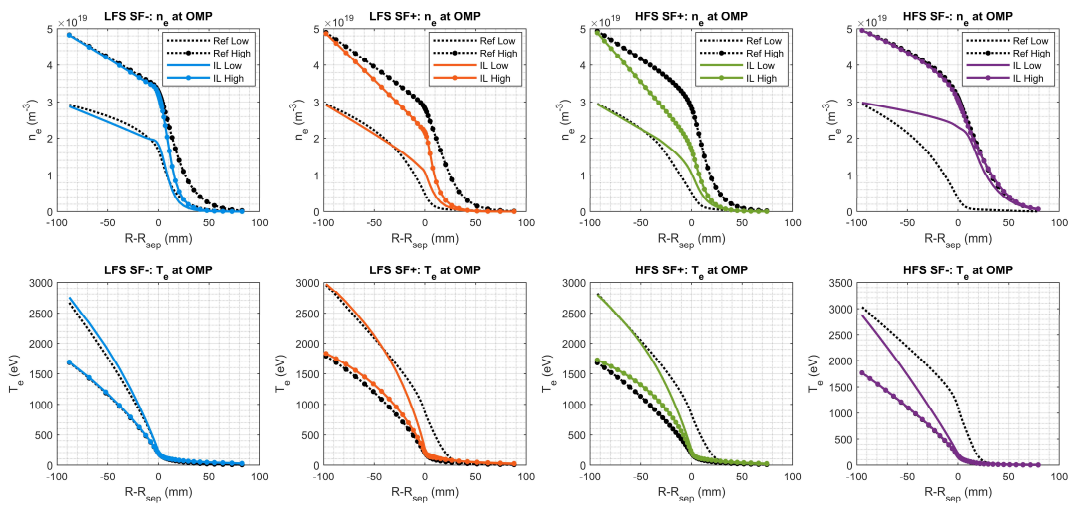

*Figure 20 OMP profiles for ITER-like (IL) cases. From left to right are LFS SF- cases, LFS SF+ cases, HFS SF+ cases and HFS SF- cases. The top row shows the electron density profiles, while the bottom row depicts the electron temperature profiles. For the low upstream density cases, the $n_{core}$ is $3\times10^{19}$ m$^{-3}$ and for the high upstream density cases, the $n_{core}$ is $5\times10^{19}$ m$^{-3}$.*

## 6.1. The LFS SF- ITER-like divertor

For the LFS SF- IL Low density and High density cases, the target profiles together with the reference cases are shown in Figure 21. The 2D distribution of neutral particle density $n_{D+2D2}$ (including atomic and molecular D$_2$), neutral pressure $Pr_{D+2D2}$ (atomic and molecular contributions), particle source $S_{D+}$ and particle loss $L_{D+}$ are shown in Figure 22.

The IL Low density case has similar upstream conditions as the Reference Low case: $n_{e,sep}\sim1.8\times10^{19}$ m$^{-3}$ and $T_{e,sep}\sim200$eV. But, the OT1 in IL Low density case is fully detached, with a peak value of $q_{\parallel}$ dropping from 300 MWm$^{-2}$ to 50 MWm$^{-2}$ resulting in $q_{surf}$ as low as 2 MWm$^{-2}$ as shown in Figure 21(j)(n). This is attributed to the closed target structure with high inclined plates, especially the inner target. With the IL divertor geometry, recycling neutrals from IT1 are compressed to the PFR1 region. Compared to the Reference Low case, even though the upstream conditions are very similar, the OT1 is already fully detached thanks to the high volumetric dissipation, similarly to the Reference High case. A strong recombination region together with high neutral density and pressure visible in Figure 22 (i)-(l) confirms this explanation. A schematic view of this characteristic is plot in Figure 23.

However, a side effect is that $q_{\parallel}$ at IT1 increases compared to the reference, since the recycling neutrals are no longer trapped in the HFS SOL region. This effect can be mitigated by the inclined target plate, which reduces the poloidal tilting angle α. As a result, the peak value of $q_{surf}$ at IT1 in the IL low density case is only 50% of that in the reference case as in Figure 21 (m). In the IL High density case, IT1, OT1, and OT2 exhibit the same power exhaust performance as in the IL Low density case, with similar peak $q_{surf}$ values. The total recombination rates for IL Low density and High cases are $6.51\times10^{23}$ s-1 and $5.88\times10^{23}$ s$^{-1}$. In this work, the pumping is below PFR1. It is speculated that if the pumping surface is moved to the PFR2 region, $q_{surf}$ at OT2 in the IL High density case can be further decreased, because recycling neutrals from OT2 tend to remain longer in the divertor region before being pumped out, which facilitates volumetric dissipation.

By integrating the results from this section with the analyses presented in Sections 4 and 5, a new strategy for designing the LFS SF- divertor is proposed:

1. **Placement of the secondary X-point**: The introduction of the secondary X-point splits the SOL into SOL1 and SOL2, effectively "bending" LFS SOL1 toward the HFS side. Therefore, the secondary X-point needs to be positioned so as to minimize the distance between IT1 and OT1, bringing the two targets into the closest possible proximity.
2. **IT1 target shaping**: The shape of the IT1 target should compress recycling neutrals, efficiently directing them into the PFR3 and LFS SOL1 regions to enhance volumetric energy dissipation. As a result, the OT1 target can benefit from this dissipation and more readily achieve detachment.

3. **Inclined target plates**: Highly inclined target plates for IT1 and OT2 are recommended, as they can significantly reduce the surface heat flux $q_{surf}$ through reducing the poloidal tilting angle.
4. **Optimization of the distance *dxx*:** On the one hand, *dxx* should be optimized to fully utilize volumetric dissipation and remove as much power as possible in the SOL1 region. On the other hand, it should allow the power entering SOL2 to be effectively mitigated by the inclined target angle. By balancing these two effects, the surface heat flux on both targets can be made comparable, leading to an optimized divertor design. In this study, a distance of *dxx* ≈ 3 mm ≈ ½ $\lambda_q$ appears to be a suitable choice, given a local $\lambda_q$ of approximately 6 mm.

This strategy focuses on power exhaust and does not consider impurity screening or helium pumping. In future studies, these two aspects will be considered to more comprehensively evaluate the performance of the LFS SF- divertor.

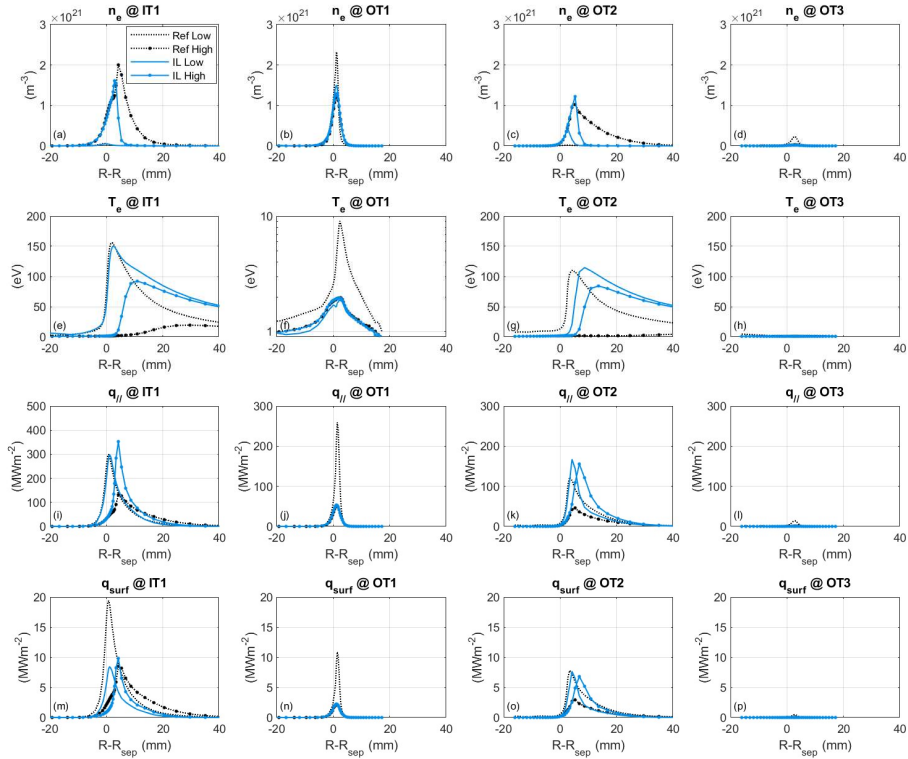

*Figure 21 Target profiles for LFS SF- reference cases (black) and LFS SF- ITER-like cases (blue). From top to bottom: electron density $n_e$, electron temperature $T_e$, parallel heat load $q_{\parallel}$ and target heat load $q_{surf}$. From left to right: profiles at IT1, OT1, OT2, OT3. The values at the OT3 are negligible.*

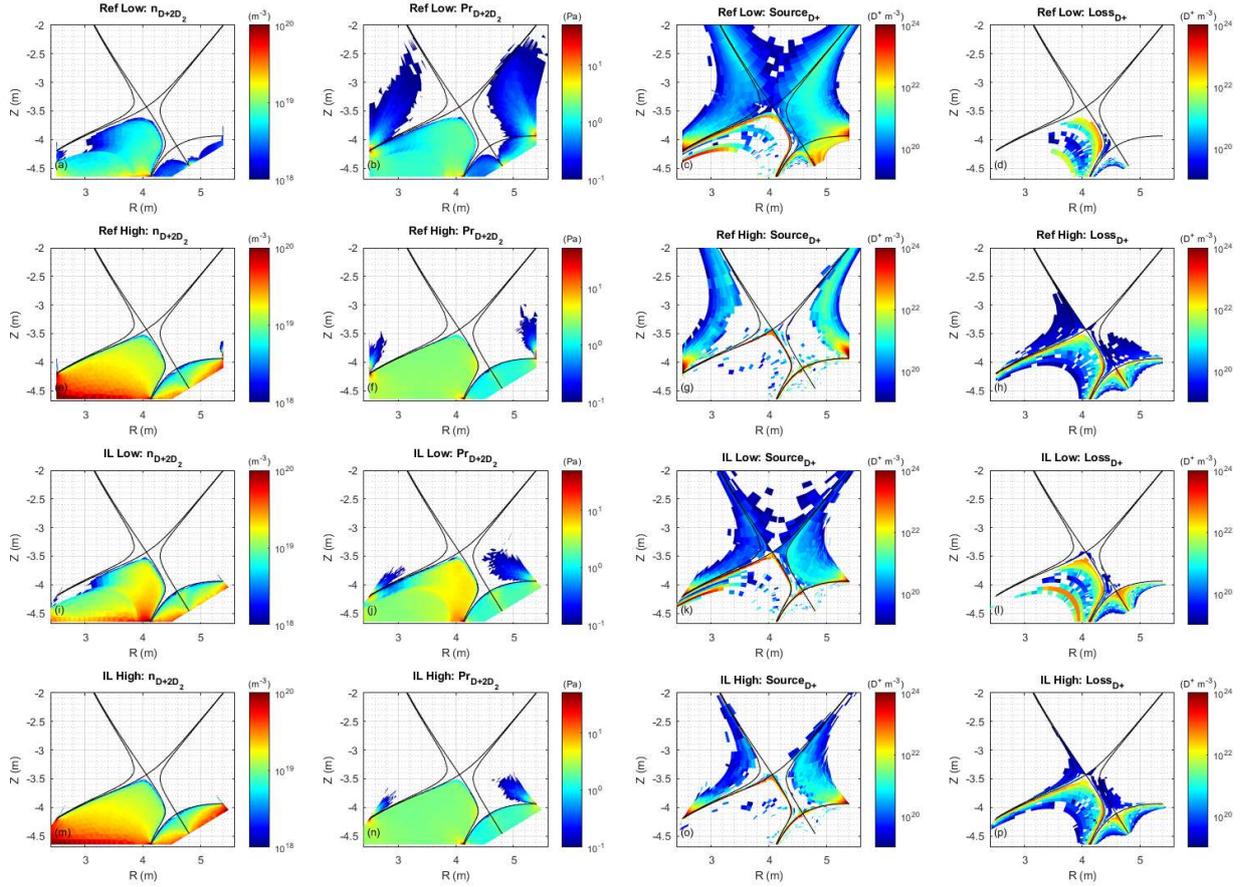

*Figure 22 2D distributions of LFS SF- reference cases and LFS SF- ITER-like cases. From top to bottom: reference low density, reference high density, ITER-like Low density case and ITER-like High density case. From left to right: neutral density $n_{D+2D2}$, neutral pressure $Pr_{D+2D2}$, particle source $S_{D+}$ and particle loss $L_{D+}$ due to recombination.*

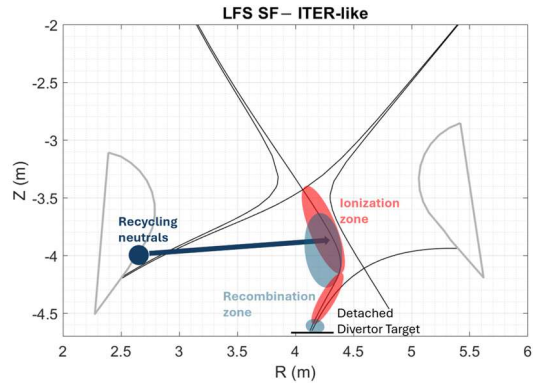

*Figure 23 A schematic view of the LFS SF- characteristic that recycling neutrals is compressed to LFS SOL region results in the detachment of outer target.*

## 6.2. The LFS SF+ ITER-like divertor

For the LFS SF+ IL cases, the target profiles together with the reference cases are shown in Figure 24 and the 2D distribution are shown in Figure 25. With the IL divertor geometry, $q_\parallel$ at IT1 and OT1 increases in the IL Low density and IL High density cases. This is similar to the IT1 of LFS SF- IL cases, where the recycling neutrals are baffled to the four PFRs instead of being trapped in the SOL. However, the IL divertor geometry has smaller poloidal tilting angle resulting in the same $q_{surf}$ values compared to the reference cases. For IT1, the peak value of $q_{surf}$ is

~10MWm$^{-2}$ and for OT1 the peak value of $q_{surf}$ is ~15MWm$^{-2}$ as shown in Figure 24 (m)(n). From Figure 25, compared to the reference cases, the neutral pressure in the PFRs increases and a high neutral density region is formed around the PFR1 region in the IL cases. Ionization and recombination zones are observed in IL Low density and High density cases in Figure 25(k)(l)(o)(p).

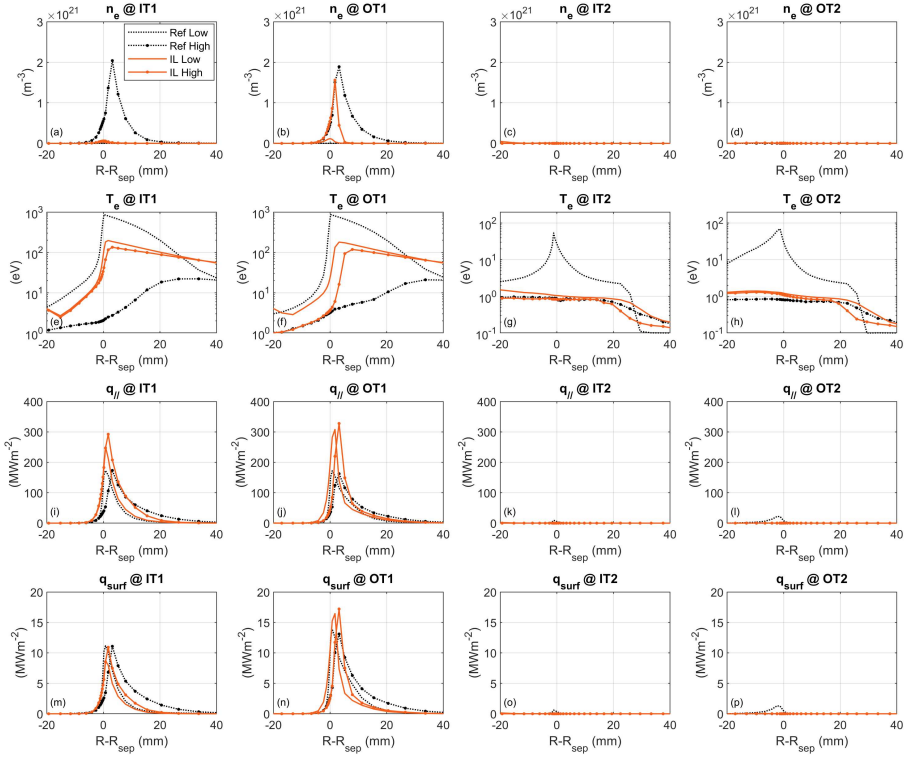

*Figure 24 Target profiles for LFS SF+ reference cases (black) and LFS SF+ ITER-like cases (orange). From top to bottom: electron density $n_e$, electron temperature $T_e$, parallel heat load $q_{\parallel}$ and target heat load $q_{surf}$. From left to right: profiles at IT1, OT1, IT2, OT2.*

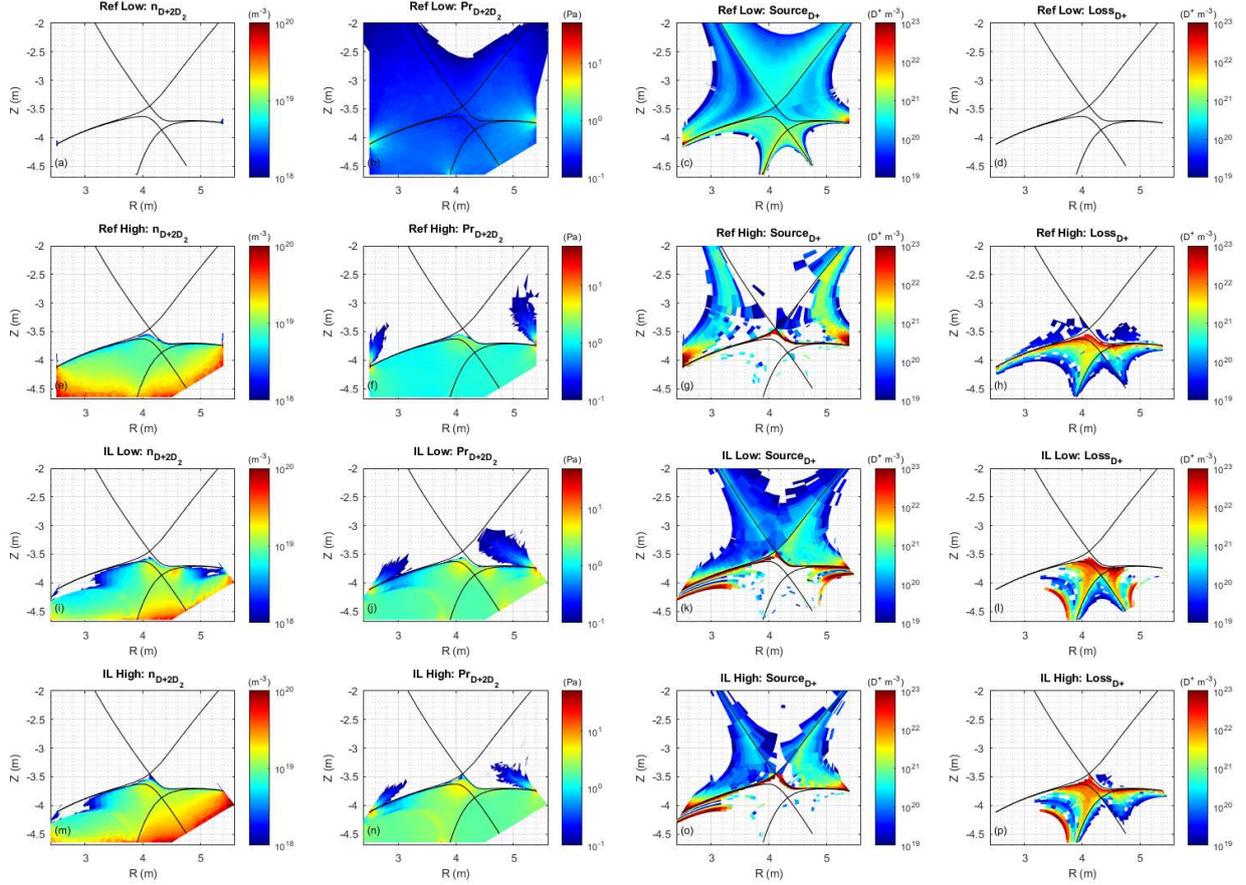

*Figure 25 2D distributions of LFS SF+ reference cases and LFS SF- ITER-like cases. From the top to bottom: Reference Low density case, Reference High density case, ITER-like Low density case and ITER-like High density case. From left to right: neutral density $n_{D+2D2}$, neutral pressure $Pr_{D+2D2}$, particle source $S_{D+}$ and particle loss $L_{D+}$ due to recombination.*

### 6.3. The HFS SF+ ITER-like divertor

For the HFS SF+ IL cases, the target profiles together with the reference cases are shown in Figure 26 and the 2D distribution are reported in Figure 27. Compared to the reference case, at IT1, $q_{\parallel}$ in the IL Low density case increases but $q_{surf}$ stays at the same level. At OT1, $q_{\parallel}$ in the IL Low density case is comparable with the reference case but the $q_{surf}$ decreases by approximately 50% due to the titling angle. There is a high neutral density region form within the PFR1 region near the primary X-point in Figure 27. Compared to LFS SF+ cases, we believe that this is due to $\theta$. In fact, as mentioned in section 2, $\theta$ in HFS SF+ is close to 90° while in LFS SF+ it is ~52°. There is a trade-off between the PFR1 volume and the connection length in PFR1: when the two X-points are close to each other, the connection length becomes large, while the PFR1 volume reduces, and vice versa. It is speculated that when $\theta$ approaches 90°, both the PFR1 volume and the connection length may be favorably balanced, which could facilitate the formation of a high neutral density region. This suggests that a scan of $\theta$ could be performed in future study. For both LFS and HFS SF+ IL cases, around the primary X-point, there is strong ionization source and recombination source in the PFR1 region. These features may be beneficial for the X-point Radiator, which exhibits strong ionization and recombination source in the core region near the X-point [39][40], but future simulations with impurity seeding are necessary.

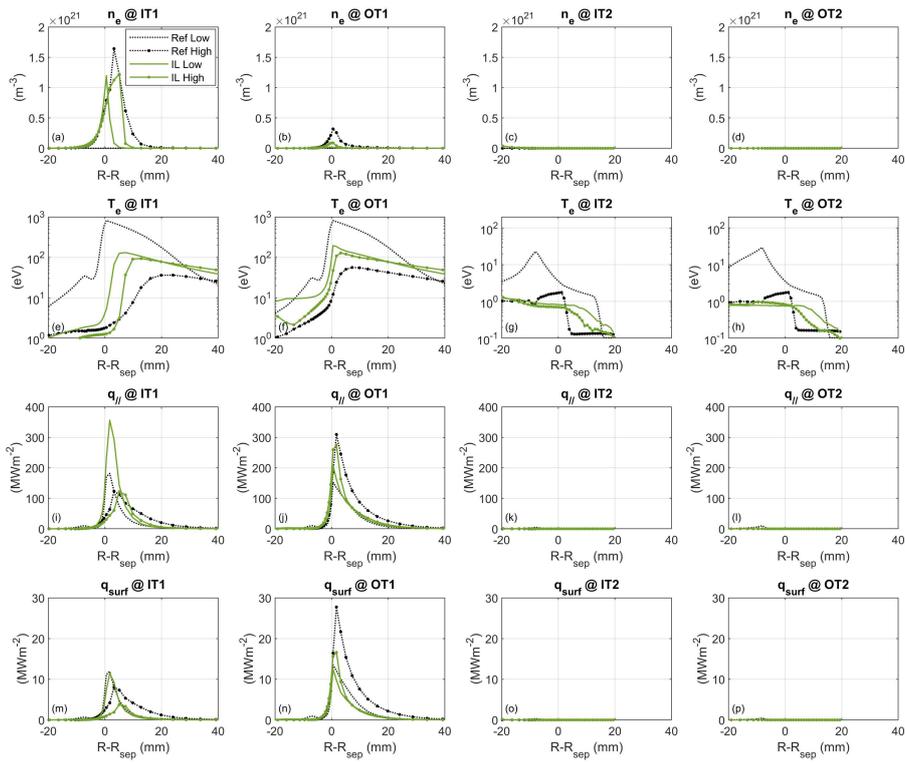

*Figure 26 Target profiles for LFS SF+ reference cases (black) and LFS SF+ ITER-like cases (green). From top to bottom: electron density $n_e$, electron temperature $T_e$, parallel heat load $q_\parallel$ and target heat load $q_{surf}$. From left to right: profiles at IT1, OT1, IT2, OT2.*

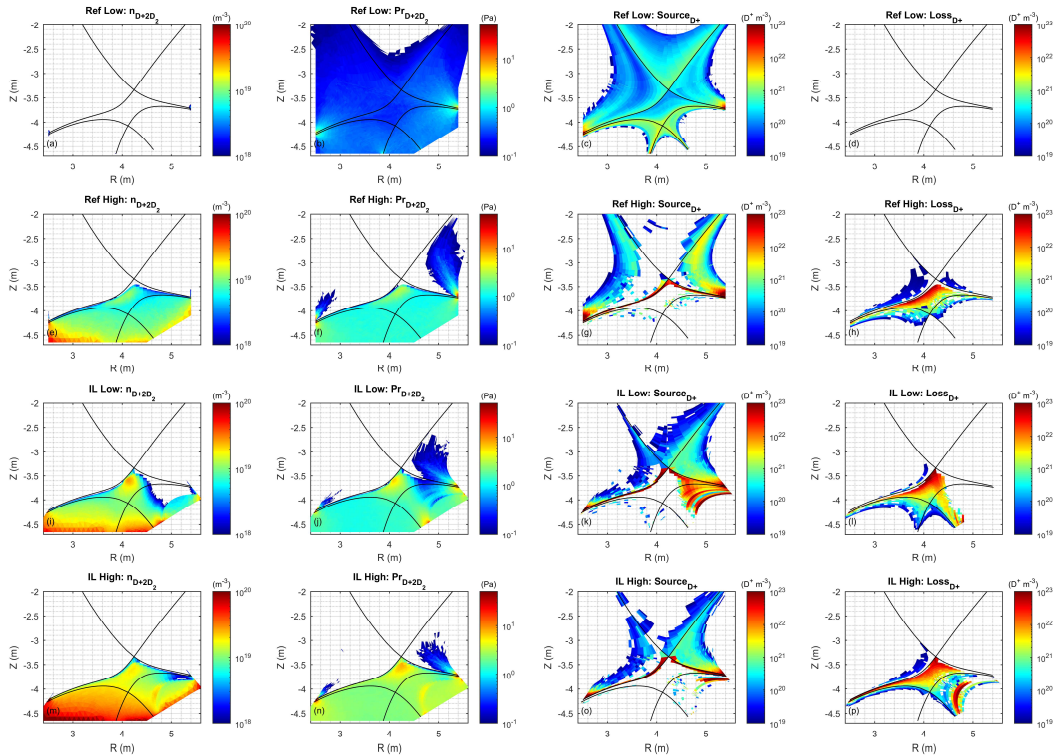

*Figure 27 2D distributions of LFS SF+ reference cases and LFS SF- ITER-like cases. From the top to bottom: Reference Low density case, Reference High density case, ITER-like Low density and ITER-like High density case. From left to right: neutral density $n_{D+2D2}$, neutral pressure $Pr_{D+2D2}$, particle source $S_{D+}$ and particle loss $L_{D+}$ due to recombination.*

## 6.4. The HFS SF- ITER-like divertor

For the HFS SF- IL cases, the target profiles together with the reference cases are reported in Figure 28 and the 2D distributions are shown in Figure 29. For the IL Low density case, the IL geometry results in peak value $q_\parallel$ reducing from 180 MWm$^{-2}$ to 130 MWm$^{-2}$ at IT1 and from 100 MWm$^{-2}$ to 50 MWm$^{-2}$ at IT2. This is because the inner IL divertor plate compresses the recycling neutrals to SOL2, so favoring strong volumetric dissipation. For the OT1 target, the IL divertor structure works as a neutral baffling without changing the poloidal titling angle as the separatrix is not terminated on the IL divertor structure. In our simulation, the SOL mesh is wide. Even with high neutral pressure and density, the neutral particles cannot penetrate in the SOL deeply. This can be confirmed by Figure 29 (h) and (p) that the ionization distribution near the OT1 is the same level. Thus, the peak values of $q_\parallel$ and $q_{surf}$ at OT1 are the same between reference and IL cases as shown in Figure 28(k)(o). These simulations suggest that in future study about the HFS SF- IL divertor plate at LFS should be optimized to intersect the separatrix line, e.g. by shifting the target plate toward the HFS. This optimization is expected to compress the recycling neutrals from OT1 in the PFR1 and HFS SOL1 regions, thereby further reducing $q_\parallel$ at OT1. In principle, the HFS SF- configuration can be combined with the Super-X outer divertor; we plan to explore this possibility in the future.

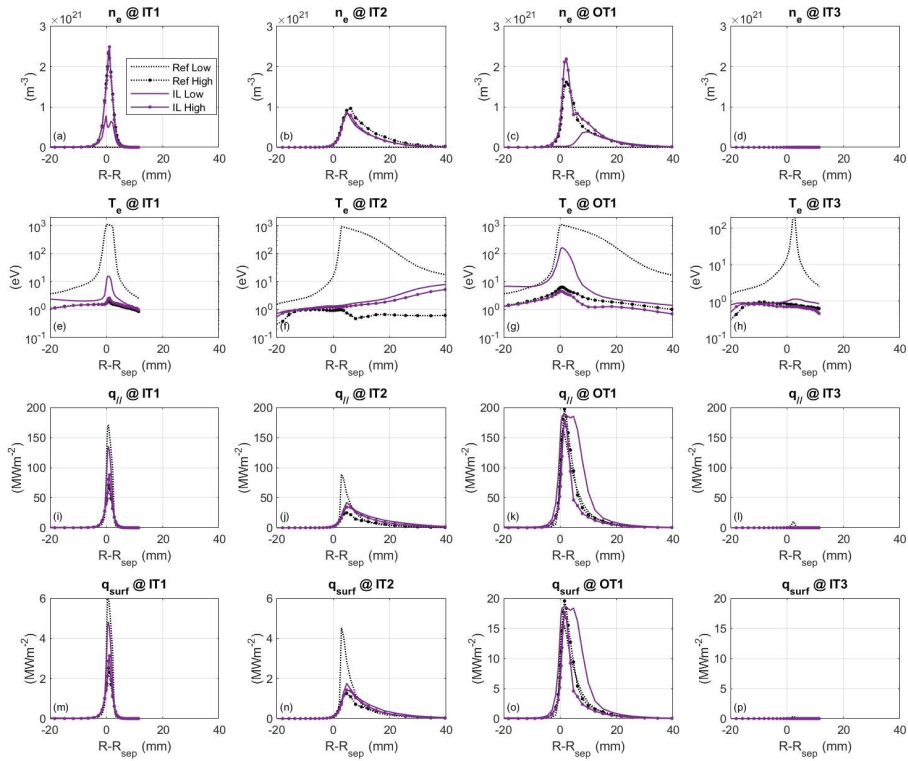

*Figure 28 Target profiles for HFS SF- reference cases (black) and LFS SF+ ITER-like cases (purple). From top to bottom: electron density $n_e$, electron temperature $T_e$, parallel heat load $q_\parallel$ and target heat load $q_{surf}$. From left to right: profiles at IT1, IT2, OT1, IT3.*

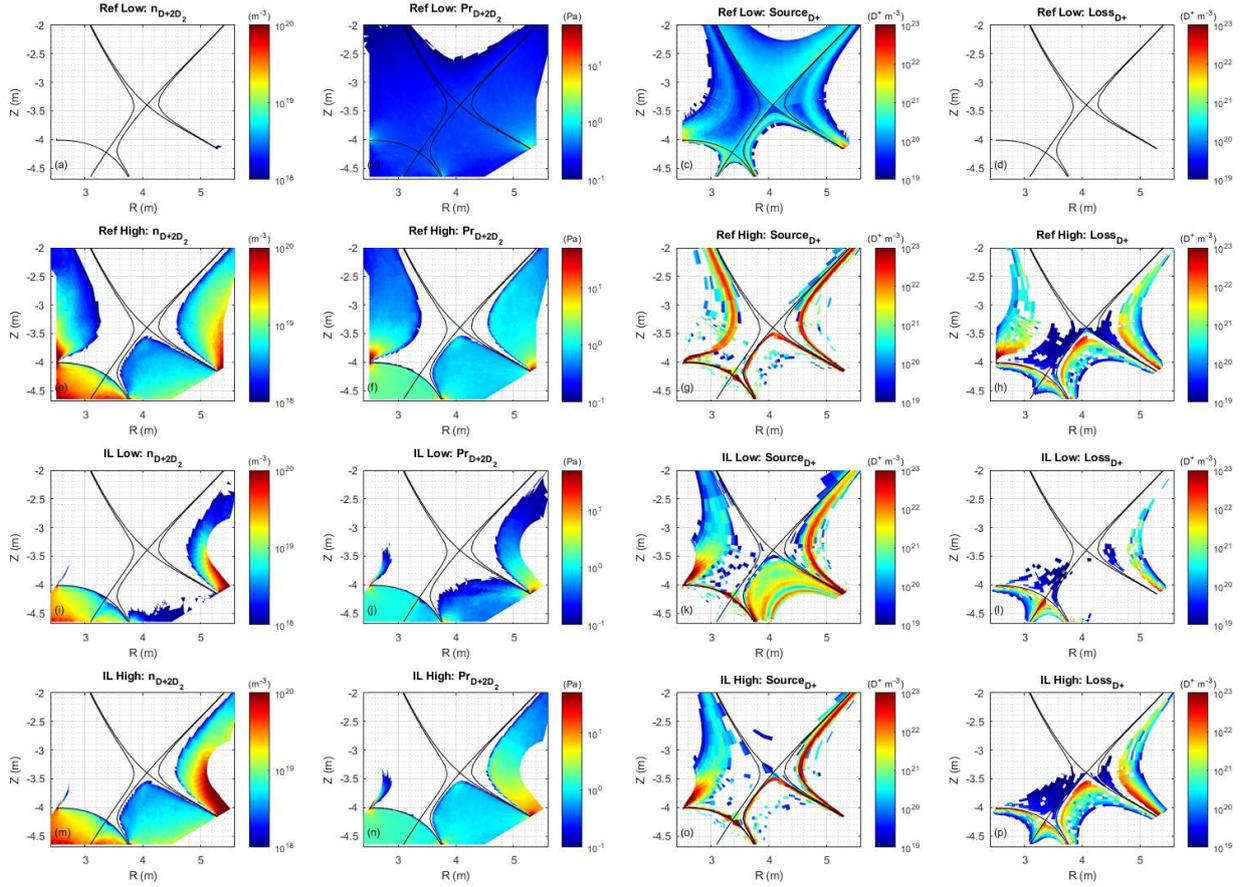

*Figure 29 2D distributions of HFS SF- reference cases and LFS SF- ITER-like cases. From top to bottom: Reference Low density case, Reference High density case, ITER-like Low density case and ITER-like High density case. From left to right: neutral density $n_{D+2D2}$, neutral pressure $Pr_{D+2D2}$, particle source $S_{D+}$ and particle loss $L_{D+}$ due to recombination.*

# 7. Summary and Outlook

In this paper, ITER-scale snowflake (SF) divertors are numerically studied with the SOLPS-ITER code. Four configuration types are considered including: Low-Field Side SF− (LFS SF−), Low-Field Side SF+ (LFS SF+), High-Field Side SF+ (HFS SF+), and High-Field Side SF− (HFS SF−). An upstream density scan is performed so as to span typical divertor regimes: low-recycling, high-recycling, and detachment. The secondary X-point positions are varied in order to examine the influence of the magnetic geometry in detail. Specifically, for LFS SF− and HFS SF−, the scan on dxx (the distance between the two X-points measured at the outer midplane) is carried out. For LFS SF+ and HFS SF+, a scan of σ (the normalized distance between the two X-points) is performed. The trend of power splitting at the secondary X-point is consistent with experimental observations, whereas the in–out power sharing remains nearly constant, which is inconsistent with experimental results. Finally, the effect of divertor geometry is assessed by comparing results from simplified flat plate targets with those from ITER-like target geometries.

According to the overall simulation results, there is a noticeable consequence of LFS SF- divertor: a closed divertor structure with inclined target plates can effectively compress recycling neutrals originating from the HFS divertor region into the LFS SOL and PFR regions, leading to strong volumetric power dissipation through ionization and recombination in the SOL region. As a result, the outer target which is magnetically connected to the dissipation region is easily detached.

Based on these findings, a possible strategy for designing the LFS SF− divertor is proposed:
- The placement of the secondary X-point should minimize the distance between the IT1 and OT1 targets.
- The shaping of the IT1 target should promote compression of recycling neutrals from the inner divertor region into the LFS SOL region.
- The target plates should be inclined to produce a smaller poloidal tilting angle, thereby reducing the target heat load.
- The parameter *dxx* can be used to control power exhaust performance across all targets, enabling balanced heat load behavior.

For the LFS SF+ and HFS SF+ divertors, a high-density region between the two X-points, associated with both ionization and recombination zones, is observed. With the ITER-like divertor geometry, a high neutral density zone in the PFR region close to the primary X-point can be formed. This feature might be beneficial for the formation of X-point radiator but require further impurity seeding simulations.

In all the simulations of this study, uniform perpendicular transport coefficients, representing turbulent transport, are used across all plasma computational domains. In a previous AUG validation study [35], the perpendicular transport coefficients in the core and SOL regions were fine-tuned to match experimental measurements, while uniform values were retained for the PFR. In the case of SF configurations, the PFRs extends over a larger spatial domain than SN configuration. Recent TCV studies [43][44] indicate the existence of strong turbulence within the PFRs of SF divertors that redistribute the power flux. MHD simulations of MAST-U show that the diffusion coefficient is significantly enhanced in the vicinity of the X-point [45]. Therefore, applying uniform perpendicular transport coefficients for the whole PFRs may not be appropriate. Future SOLPS-ITER modeling of SF divertors should focus more on exploring the transport behavior within the PFRs e.g. enhanced cross-field transport and the drifts behavior, in order to improve confidence in predicted simulations of future devices.

In this study, the core boundary condition is prescribed as a fixed core density instead of gas puffing fueling. The simplified pumping surfaces in the four divertor configurations are not identical; although efforts were made to minimize these discrepancies, no common position could be applied to all cases. Consequently, different particle throughputs arise even under similar upstream conditions e.g. under the high upstream density, it is range from $1.9 \times 10^{21}$ to $3.5 \times 10^{21}$ atom/s for SF divertors, which in turn affect the divertor performance. In future work, gas puff fueling and impurity seeding will be considered to ensure consistent throughputs to evaluate divertor performance. Besides, the dome structure is not included. In reality, it can significantly affect the neutral dynamics in the sub-divertor region, which is closely associated with the complex PFRs in the SF divertors and should be investigated in more detail in future work to better utilize SF divertors

## Acknowledgements

Haosheng Wu is deeply grateful to Prof. Yühe Feng and Dr. Xavier Bonnin for their inspired discussions and encouragements. This work is supported by PNRR M4C2 - HPC, Big data and Quantum Computing (Simulazioni, calcolo e analisi dei dati e altre prestazioni - CN1) - CUP I53C22000690001 SPOKE 6 Multiscale modelling & Engineering applications.